\documentclass[12pt]{article}
\usepackage{a4p}
\usepackage{citesort}
\usepackage{pennames}
\usepackage{epsfig}

\begin{document}
\newcommand{\EE}        {\ensuremath{\mathrm{e}^+\mathrm{e}^-}} 
\newcommand{\GG}        {\ensuremath{\mathrm{\gamma\gamma}}} 
\newcommand{\GSGS}      {\ensuremath{\mathrm{\gamma^{*}\gamma^{*}}}} 
\newcommand{\PP}        {\ensuremath{\mathrm{p\bar{p}}}} 
\newcommand{\LL}        {\ensuremath{\mathrm{\Lambda\overline{\Lambda}}}} 
\newcommand{\XIXI}      {\ensuremath{\mathrm{\Xi^{-}\Xi^{+}}}} 
\newcommand{\MuMu}      {\ensuremath{\mathrm{\mu^{+}\mu^{-}}}} 
\newcommand{\PiPi}      {\ensuremath{\mathrm{\pi^{+}\pi^{-}}}}
\newcommand{\KK}        {\ensuremath{\mathrm{K^{+}K^{-}}}}
\newcommand{\K}         {\ensuremath{\mathrm{K^{0}}}}
\newcommand{\TT}        {\ensuremath{\mathrm{\tau^{+}\tau^{-}}}}
\newcommand{\KsKs}      {\ensuremath{\mathrm{K^0_S K^0_S}}}
\newcommand{\X}         {\ensuremath{\mathrm{X}}}
\newcommand{\m}         {\rm m}
\newcommand{\cm}        {\rm cm}
\newcommand{\mm}        {\rm mm}
\newcommand{\mrad}      {\rm mrad}
\newcommand{\TP}        {\ensuremath{\mathrm{{\theta}-{\phi}}}}
\newcommand{\EV}        {\ensuremath{\mathrm{eV}}}
\newcommand{\KV}        {\ensuremath{\mathrm{keV}}}
\newcommand{\GV}        {\ensuremath{\mathrm{GeV}}}
\newcommand{\GVc}       {\ensuremath{\mathrm{GeV/c}}}
\newcommand{\MV}        {\ensuremath{\mathrm{MeV}}}
\newcommand{\SI}        {\ensuremath{\mathrm{\sigma}}}
\newcommand{\costs}     {\ensuremath{\mathrm{|\cos\theta^{*}|}}}
\newcommand{\cost}      {\ensuremath{\mathrm{|\cos\theta|}}}
\newcommand{\costsgen}  {\ensuremath{\mathrm{|\cos\theta^{*}_{\rm GEN}|}}}
\newcommand{\spts}      {\ensuremath{|\sum{\vec{p}_\perp}|^{2}}}
\newcommand{\ETTo}      {\ensuremath{\mathrm{\varepsilon}_{\rm TT1}}}
\newcommand{\ETTt}      {\ensuremath{\mathrm{\varepsilon}_{\rm TT2}}}
\newcommand{\ETOo}      {\ensuremath{\mathrm{\varepsilon}_{\rm TOF1}}}
\newcommand{\ETOt}      {\ensuremath{\mathrm{\varepsilon}_{\rm TOF2}}}
\newcommand{\EEMo}      {\ensuremath{\mathrm{\varepsilon}_{em1}}}
\newcommand{\EEt}      {\ensuremath{\mathrm{\varepsilon}_{em2}}}
\newcommand{\T}         {\ensuremath{\mathrm{\theta}}}
\newcommand{\LUM}       {\ensuremath{L}}
\newcommand{\NE}        {\ensuremath{N_{\rm ev}}}
\newcommand{\EDET}      {\ensuremath{\mathrm{\varepsilon}_{\rm DET}}}
\newcommand{\ETRIG}     {\ensuremath{\mathrm{\varepsilon}_{\rm TRIG}}}
\newcommand{\SQS}       {\ensuremath{\sqrt{s_{\rm ee}}}}
\newcommand{\NS}        {\ensuremath{N_{\rm SEL}}}
\newcommand{\NGEN}      {\ensuremath{N_{\rm GEN}}}
\newcommand{\D}         {\ensuremath{\mathrm{\bigtriangleup}}}
\newcommand{\PM}        {\ensuremath{\mathrm{\pm}}}
\newcommand{\PT}        {\ensuremath{p_{\perp}}}
\newcommand{\dedx}      {\ensuremath{{\rm d}E/{\rm d}x}}
\newcommand{\DEDX}      {\ensuremath{{\rm d}E/{\rm d}x}}
\newcommand{\lumi}      {\ensuremath{\mathcal{L}}}
\newcommand{\cosw}      {\rule{0pt}{14pt}\lower0.5em\hbox{$W(\GV)$}$\backslash$\raise0.5em\hbox{$\costs$}}
\newcommand{\mb}        {\ensuremath{\mathrm{mb}}}
\newcommand{\nb}        {\ensuremath{\mathrm{nb}}}
\newcommand{\pb}        {\ensuremath{\mathrm{pb}}}
\def\pz{\phantom{0}}
%
%
%
%
\newcommand{\Wa}        {\ensuremath{2.1}}
\newcommand{\Wb}        {\ensuremath{2.2}}
\newcommand{\Wc}        {\ensuremath{2.3}}
\newcommand{\Wd}        {\ensuremath{2.4}}
\newcommand{\We}        {\ensuremath{2.5}}
\newcommand{\Wf}        {\ensuremath{2.6}}
\newcommand{\Wg}        {\ensuremath{2.7}}
\newcommand{\Wh}        {\ensuremath{2.8}}
\newcommand{\Wl}        {\ensuremath{2.9}}
\newcommand{\Wm}        {\ensuremath{3.0}}
\newcommand{\Wn}        {\ensuremath{3.1}}
\newcommand{\Wo}        {\ensuremath{3.2}}
\newcommand{\Wp}        {\ensuremath{3.3}}
\newcommand{\Wq}        {\ensuremath{3.4}}
\newcommand{\Wr}        {\ensuremath{3.5}}
\newcommand{\Ws}        {\ensuremath{3.6}}
\newcommand{\Wt}        {\ensuremath{3.7}}
\newcommand{\Wu}        {\ensuremath{3.8}}
\newcommand{\Wv}        {\ensuremath{3.9}}
\newcommand{\Ww}        {\ensuremath{4.0}}
\newcommand{\MWb}      {\ensuremath{2.20}}
\newcommand{\MWc}      {\ensuremath{2.30}}
\newcommand{\MWd}      {\ensuremath{2.40}}
\newcommand{\MWe}      {\ensuremath{2.50}}
\newcommand{\MWfg}     {\ensuremath{2.65}}
\newcommand{\MWh}      {\ensuremath{2.80}}
\newcommand{\MWhl}     {\ensuremath{2.85}}
\newcommand{\MWlo}     {\ensuremath{3.14}}
\newcommand{\MWpv}     {\ensuremath{3.64}}

%
%
%
%
\newcommand{\Wab}       {\ensuremath{2.15}}
\newcommand{\Wac}       {\ensuremath{2.25}}
\newcommand{\Wad}       {\ensuremath{2.35}}
\newcommand{\Wae}       {\ensuremath{2.45}}
\newcommand{\Waf}       {\ensuremath{2.55}}
\newcommand{\Wag}       {\ensuremath{2.65}}
\newcommand{\Wah}       {\ensuremath{2.75}}
\newcommand{\Wal}       {\ensuremath{2.85}}
\newcommand{\Wam}       {\ensuremath{2.95}}
\newcommand{\Wan}       {\ensuremath{3.05}}
\newcommand{\Wao}       {\ensuremath{3.15}}
\newcommand{\Wap}       {\ensuremath{3.25}}
\newcommand{\Waq}       {\ensuremath{3.35}}
\newcommand{\War}       {\ensuremath{3.45}}
\newcommand{\Was}       {\ensuremath{3.55}}
\newcommand{\Wat}       {\ensuremath{3.65}}
\newcommand{\Wau}       {\ensuremath{3.75}}
\newcommand{\Wav}       {\ensuremath{3.85}}
\newcommand{\Waz}       {\ensuremath{3.95}}
\newcommand{\Naa}       {\ensuremath{-}}                    
\newcommand{\Nab}       {\ensuremath{52}}                   
\newcommand{\Nac}       {\ensuremath{33}}                   
\newcommand{\Nad}       {\ensuremath{32}}                   
\newcommand{\Nae}       {\ensuremath{20}}                   
\newcommand{\Naf}       {\ensuremath{8}}                    
\newcommand{\Nag}       {\ensuremath{11}}                   
\newcommand{\Nafg}      {\ensuremath{18}}                   
\newcommand{\Nah}       {\ensuremath{\pz3}}                 
\newcommand{\Nal}       {\ensuremath{\pz4}}                 
\newcommand{\Nahl}      {\ensuremath{\pz6}}                 
\newcommand{\Namo}      {\ensuremath{\pz1}}                 
\newcommand{\Napv}      {\ensuremath{\pz1}}                 
%
%
\newcommand{\Snea}       {\ensuremath{-}}                   
\newcommand{\Sneb}       {\ensuremath{2.69\pm0.39\pm0.28}} 
\newcommand{\Snec}       {\ensuremath{1.53\pm0.27\pm0.16}} 
\newcommand{\Sned}       {\ensuremath{1.39\pm0.26\pm0.14}} 
\newcommand{\Snee}       {\ensuremath{0.96\pm0.22\pm0.10}} 
\newcommand{\Snefg}      {\ensuremath{0.62\pm0.22\pm0.08}} 
\newcommand{\Snehl}      {\ensuremath{0.19\pm0.11\pm0.02}} 
\newcommand{\Snemo}      {\ensuremath{0.05\,^{+0.11}_{-0.04}\pm0.01}}
\newcommand{\Snepv}      {\ensuremath{0.05\,^{+0.11}_{-0.04}\pm0.01}}
\newcommand{\Seea}       {\ensuremath{-}}              
\newcommand{\Seeb}       {\ensuremath{5.46\pm0.76}}    
\newcommand{\Seec}       {\ensuremath{2.95\pm0.51}}    
\newcommand{\Seed}       {\ensuremath{2.53\pm0.45}}              
\newcommand{\Seee}       {\ensuremath{1.69\pm0.38}}              
\newcommand{\Seef}       {\ensuremath{0.80}}              
\newcommand{\Seeg}       {\ensuremath{1.18}}              
\newcommand{\Seeh}       {\ensuremath{0.28}}              
\newcommand{\Seel}       {\ensuremath{0.29}}              
\newcommand{\Seefg}      {\ensuremath{1.01\pm0.24}}    
\newcommand{\Seehl}      {\ensuremath{0.28\pm0.12}}              
\newcommand{\Seemo}      {\ensuremath{0.26\pm0.26}}    
\newcommand{\Seepv}      {\ensuremath{0.28\pm0.28}}              
%
%
%
%
\newcommand{\defao}       {\ensuremath{0.053\pm0.005}}
\newcommand{\defat}       {\ensuremath{0.050\pm0.005}}
\newcommand{\defatr}      {\ensuremath{0.045\pm0.005}}
\newcommand{\defaf}       {\ensuremath{0.038\pm0.004}}
\newcommand{\defafi}      {\ensuremath{0.022\pm0.003}}
\newcommand{\defas}       {\ensuremath{0.013\pm0.003}}
%
%
\newcommand{\defbo}       {\ensuremath{0.053\pm0.005}}     
\newcommand{\defbt}       {\ensuremath{0.050\pm0.005}}
\newcommand{\defbtr}      {\ensuremath{0.045\pm0.005}}
\newcommand{\defbf}       {\ensuremath{0.038\pm0.004}}
\newcommand{\defbfi}      {\ensuremath{0.022\pm0.003}}
\newcommand{\defbs}       {\ensuremath{0.013\pm0.003}}
%
%
\newcommand{\defco}       {\ensuremath{0.062\pm0.005}}     
\newcommand{\defct}       {\ensuremath{0.065\pm0.006}}
\newcommand{\defctr}      {\ensuremath{0.048\pm0.005}}
\newcommand{\defcf}       {\ensuremath{0.037\pm0.004}}
\newcommand{\defcfi}      {\ensuremath{0.033\pm0.004}}        
\newcommand{\defcs}       {\ensuremath{0.028\pm0.004}}
%
%
\newcommand{\defdo}       {\ensuremath{0.070\pm0.006}}      
\newcommand{\defdt}       {\ensuremath{0.068\pm0.006}}
\newcommand{\defdtr}      {\ensuremath{0.052\pm0.005}}
\newcommand{\defdf}       {\ensuremath{0.049\pm0.005}}
\newcommand{\defdfi}      {\ensuremath{0.041\pm0.004}}        
\newcommand{\defds}       {\ensuremath{0.028\pm0.004}}
%
%
\newcommand{\deffo}       {\ensuremath{0.065\pm0.006}}     
\newcommand{\defft}       {\ensuremath{0.066\pm0.006}}
\newcommand{\defftr}      {\ensuremath{0.062\pm0.005}}
\newcommand{\defff}       {\ensuremath{0.053\pm0.005}}
\newcommand{\defffi}      {\ensuremath{0.045\pm0.005}}
\newcommand{\deffs}       {\ensuremath{0.023\pm0.003}}
%
%
\newcommand{\defgo}       {\ensuremath{0.064\pm0.006}}     
\newcommand{\defgt}       {\ensuremath{0.055\pm0.005}}
\newcommand{\defgtr}      {\ensuremath{0.062\pm0.005}}    
\newcommand{\defgf}       {\ensuremath{0.044\pm0.005}}
\newcommand{\defgfi}      {\ensuremath{0.042\pm0.005}}
\newcommand{\defgs}       {\ensuremath{0.024\pm0.003}}
%
%
\newcommand{\defho}       {\ensuremath{0.053\pm0.005}}     
\newcommand{\defht}       {\ensuremath{0.046\pm0.005}}
\newcommand{\defhtr}      {\ensuremath{0.053\pm0.005}}
\newcommand{\defhf}       {\ensuremath{0.037\pm0.004}}
\newcommand{\defhfi}      {\ensuremath{0.042\pm0.005}}
\newcommand{\defhs}       {\ensuremath{0.025\pm0.003}}
%
%
\newcommand{\deflo}       {\ensuremath{0.050\pm0.005}}     
\newcommand{\deflt}       {\ensuremath{0.043\pm0.005}}
\newcommand{\defltr}      {\ensuremath{0.040\pm0.004}}
\newcommand{\deflf}       {\ensuremath{0.033\pm0.004}}
\newcommand{\deflfi}      {\ensuremath{0.039\pm0.004}}
\newcommand{\defls}       {\ensuremath{0.025\pm0.003}}
%
%
\newcommand{\defmo}       {\ensuremath{0.047\pm0.005}}   
\newcommand{\defmt}       {\ensuremath{0.043\pm0.005}}
\newcommand{\defmtr}      {\ensuremath{0.037\pm0.004}}
\newcommand{\defmf}       {\ensuremath{0.030\pm0.004}}
\newcommand{\defmfi}      {\ensuremath{0.027\pm0.004}}
\newcommand{\defms}       {\ensuremath{0.019\pm0.003}}
%
%
\newcommand{\defno}       {\ensuremath{0.027\pm0.004}}      
\newcommand{\defnt}       {\ensuremath{0.018\pm0.003}}
\newcommand{\defntr}      {\ensuremath{0.014\pm0.003}}
\newcommand{\defnf}       {\ensuremath{0.016\pm0.003}}
\newcommand{\defnfi}      {\ensuremath{0.016\pm0.003}}
\newcommand{\defns}       {\ensuremath{0.010\pm0.002}}
%
%
%
%
\newcommand{\nevao}       {\ensuremath{20}}
\newcommand{\nevat}       {\ensuremath{16}}
\newcommand{\nevatr}      {\ensuremath{8}}
\newcommand{\nevaf}       {\ensuremath{5}}
\newcommand{\nevafi}      {\ensuremath{3}}
\newcommand{\nevas}       {\ensuremath{0}}
%
%
\newcommand{\nevbo}       {\ensuremath{10}}     
\newcommand{\nevbt}       {\ensuremath{8}}
\newcommand{\nevbtr}      {\ensuremath{9}}
\newcommand{\nevbf}       {\ensuremath{5}}
\newcommand{\nevbfi}      {\ensuremath{0}}
\newcommand{\nevbs}       {\ensuremath{1}}
%
%
\newcommand{\nevco}       {\ensuremath{12}}     
\newcommand{\nevct}       {\ensuremath{15}}
\newcommand{\nevctr}      {\ensuremath{2}}
\newcommand{\nevcf}       {\ensuremath{0}}
\newcommand{\nevcfi}      {\ensuremath{3}}        
\newcommand{\nevcs}       {\ensuremath{0}}
%
%
\newcommand{\nevdo}       {\ensuremath{7}}      
\newcommand{\nevdt}       {\ensuremath{4}}
\newcommand{\nevdtr}      {\ensuremath{1}}
\newcommand{\nevdf}       {\ensuremath{6}}
\newcommand{\nevdfi}      {\ensuremath{0}}        
\newcommand{\nevds}       {\ensuremath{2}}
%
%
\newcommand{\nevfo}       {\ensuremath{2}}     
\newcommand{\nevft}       {\ensuremath{2}}
\newcommand{\nevftr}      {\ensuremath{2}}
\newcommand{\nevff}       {\ensuremath{0}}
\newcommand{\nevffi}      {\ensuremath{0}}
\newcommand{\nevfs}       {\ensuremath{2}}
%
%
\newcommand{\nevgo}       {\ensuremath{0}}     
\newcommand{\nevgt}       {\ensuremath{1}}
\newcommand{\nevgtr}      {\ensuremath{0}}    
\newcommand{\nevgf}       {\ensuremath{2}}
\newcommand{\nevgfi}      {\ensuremath{5}}
\newcommand{\nevgs}       {\ensuremath{2}}
%
%
\newcommand{\nevho}       {\ensuremath{2}}     
\newcommand{\nevht}       {\ensuremath{1}}
\newcommand{\nevhtr}      {\ensuremath{0}}
\newcommand{\nevhf}       {\ensuremath{0}}
\newcommand{\nevhfi}      {\ensuremath{0}}
\newcommand{\nevhs}       {\ensuremath{0}}
%
%
\newcommand{\nevlo}       {\ensuremath{2}}     
\newcommand{\nevlt}       {\ensuremath{0}}
\newcommand{\nevltr}      {\ensuremath{0}}
\newcommand{\nevlf}       {\ensuremath{0}}
\newcommand{\nevlfi}      {\ensuremath{1}}
\newcommand{\nevls}       {\ensuremath{0}}
%
%
\newcommand{\nevmo}       {\ensuremath{0}}   
\newcommand{\nevmt}       {\ensuremath{0}}
\newcommand{\nevmtr}      {\ensuremath{0}}
\newcommand{\nevmf}       {\ensuremath{0}}
\newcommand{\nevmfi}      {\ensuremath{0}}
\newcommand{\nevms}       {\ensuremath{1}}
%
%
\newcommand{\nevno}       {\ensuremath{0}}      
\newcommand{\nevnt}       {\ensuremath{1}}
\newcommand{\nevntr}      {\ensuremath{0}}
\newcommand{\nevnf}       {\ensuremath{0}}
\newcommand{\nevnfi}      {\ensuremath{0}}
\newcommand{\nevns}       {\ensuremath{0}}
%
%
%
\newcommand{\bina}        {\ensuremath{0.00-0.10}}      
\newcommand{\binb}        {\ensuremath{0.10-0.20}}
\newcommand{\binc}        {\ensuremath{0.20-0.30}}
\newcommand{\bind}        {\ensuremath{0.30-0.40}}
\newcommand{\bine}        {\ensuremath{0.40-0.50}}
\newcommand{\binf}        {\ensuremath{0.50-0.60}}
%
%
%
%
\newcommand{\csao}        {\ensuremath{8.8\pm2.0}}
\newcommand{\csat}        {\ensuremath{7.4\pm1.8}}
\newcommand{\csatr}       {\ensuremath{4.4\pm1.6}}
\newcommand{\csaf}        {\ensuremath{3.0\pm1.4}}
\newcommand{\csafi}       {\ensuremath{3.2\pm1.9}}
\newcommand{\csas}        {\ensuremath{0.0{{+1.14}\atop{-0.0}}}}
%
%
\newcommand{\csbo}        {\ensuremath{3.4\pm1.1}}     
\newcommand{\csbt}        {\ensuremath{3.8\pm1.3}}
\newcommand{\csbtr}       {\ensuremath{4.4\pm1.5}}
\newcommand{\csbf}        {\ensuremath{2.7\pm1.2}}
\newcommand{\csbfi}       {\ensuremath{0.0{{+1.14}\atop{-0.0}}}}
\newcommand{\csbs}        {\ensuremath{1.0{{+2.29}\atop{-0.83}}}}
%
%
\newcommand{\csco}        {\ensuremath{4.8\pm1.4}}     
\newcommand{\csct}        {\ensuremath{5.7\pm1.5}}
\newcommand{\csctr}       {\ensuremath{1.1\pm0.8}}
\newcommand{\cscf}        {\ensuremath{0.0{{+1.14}\atop{-0.0}}}}
\newcommand{\cscfi}       {\ensuremath{2.4\pm1.4}}        
\newcommand{\cscs}        {\ensuremath{0.0{{+1.14}\atop{-0.0}}}}
%
%
\newcommand{\csdo}        {\ensuremath{2.6\pm1.0}}      
\newcommand{\csdt}        {\ensuremath{1.6\pm0.8}}
\newcommand{\csdtr}       {\ensuremath{0.5{{+1.14}\atop{-0.41}}}}
\newcommand{\csdf}        {\ensuremath{3.2\pm1.3}}
\newcommand{\csdfi}       {\ensuremath{0.0{{+1.14}\atop{-0.0}}}}     
\newcommand{\csds}        {\ensuremath{1.7\pm1.2}}
%
%
\newcommand{\csfo}        {\ensuremath{0.8\pm0.6}}     
\newcommand{\csft}        {\ensuremath{0.8\pm0.6}}
\newcommand{\csftr}       {\ensuremath{0.9\pm0.7}}
\newcommand{\csff}        {\ensuremath{0.0{{+1.14}\atop{-0.0}}}}     
\newcommand{\csffi}       {\ensuremath{0.0{{+1.14}\atop{-0.0}}}}     
\newcommand{\csfs}        {\ensuremath{2.4\pm1.7}}
%
%
\newcommand{\csgo}        {\ensuremath{0.0{{+1.14}\atop{-0.0}}}}     
\newcommand{\csgt}        {\ensuremath{0.5{{+1.14}\atop{-0.41}}}}
\newcommand{\csgtr}       {\ensuremath{0.0{{+1.14}\atop{-0.0}}}}    
\newcommand{\csgf}        {\ensuremath{1.3\pm0.9}}
\newcommand{\csgfi}       {\ensuremath{3.3\pm1.5}}
\newcommand{\csgs}        {\ensuremath{2.4\pm1.7}}
%
%
\newcommand{\csho}        {\ensuremath{1.2\pm0.6}}     
\newcommand{\csht}        {\ensuremath{0.7{{+1.60}\atop{-0.58}}}}    
\newcommand{\cshtr}       {\ensuremath{0.0{{+1.14}\atop{-0.0}}}}     
\newcommand{\cshf}        {\ensuremath{0.0{{+1.14}\atop{-0.0}}}}     
\newcommand{\cshfi}       {\ensuremath{0.0{{+1.14}\atop{-0.0}}}}
\newcommand{\cshs}        {\ensuremath{0.0{{+1.14}\atop{-0.0}}}}
%
%
\newcommand{\cslo}        {\ensuremath{1.2\pm0.9}}     
\newcommand{\cslt}        {\ensuremath{0.0{{+1.14}\atop{-0.0}}}}
\newcommand{\csltr}       {\ensuremath{0.0{{+1.14}\atop{-0.0}}}}
\newcommand{\cslf}        {\ensuremath{0.0{{+1.14}\atop{-0.0}}}}
\newcommand{\cslfi}       {\ensuremath{0.8{{+1.83}\atop{-0.66}}}}
\newcommand{\csls}        {\ensuremath{0.0{{+1.14}\atop{-0.0}}}}
%
%
\newcommand{\csmo}        {\ensuremath{0.0{{+1.14}\atop{-0.0}}}}   
\newcommand{\csmt}        {\ensuremath{0.0{{+1.14}\atop{-0.0}}}}
\newcommand{\csmtr}       {\ensuremath{0.0{{+1.14}\atop{-0.0}}}}
\newcommand{\csmf}        {\ensuremath{0.0{{+1.14}\atop{-0.0}}}}
\newcommand{\csmfi}       {\ensuremath{0.0{{+1.14}\atop{-0.0}}}}
\newcommand{\csms}        {\ensuremath{0.5{{+1.14}\atop{-0.41}}}}
%
%
\newcommand{\csno}        {\ensuremath{0.0{{+1.14}\atop{-0.0}}}}
\newcommand{\csnt}        {\ensuremath{0.5{{+1.14}\atop{-0.41}}}}
\newcommand{\csntr}       {\ensuremath{0.0{{+1.14}\atop{-0.0}}}}
\newcommand{\csnf}        {\ensuremath{0.0{{+1.14}\atop{-0.0}}}}
\newcommand{\csnfi}       {\ensuremath{0.0{{+1.14}\atop{-0.0}}}}
\newcommand{\csns}        {\ensuremath{0.0{{+1.14}\atop{-0.0}}}}
%
%
\newcommand{\csta}     {\ensuremath{0.80\pm0.33\pm0.08}}      
\newcommand{\cstb}     {\ensuremath{0.51\pm0.26\pm0.05}}
\newcommand{\cstc}     {\ensuremath{0.23\pm0.17\pm0.03}}
\newcommand{\cstd}     {\ensuremath{0.31\pm0.22\pm0.04}}
\newcommand{\cste}     {\ensuremath{1.03\pm0.42\pm0.13}}
\newcommand{\cstf}     {\ensuremath{1.17\pm0.58\pm0.16}}
%
%
%
\newcommand{\cstaaa}     {\ensuremath{4.91\pm0.70\pm0.48}}      
\newcommand{\cstbaa}     {\ensuremath{4.62\pm0.71\pm0.45}}
\newcommand{\cstcaa}     {\ensuremath{2.61\pm0.58\pm0.28}}
\newcommand{\cstdaa}     {\ensuremath{2.24\pm0.56\pm0.23}}
\newcommand{\csteaa}     {\ensuremath{1.40\pm0.58\pm0.19}}
\newcommand{\cstfaa}     {\ensuremath{0.67\pm0.39\pm0.09}}
%
%
%
\newcommand{\cstab}     {\ensuremath{1.87\pm0.39\pm0.18}}      
\newcommand{\cstbb}     {\ensuremath{1.86\pm0.39\pm0.19}}
\newcommand{\cstcb}     {\ensuremath{0.50\pm0.23\pm0.05}}
\newcommand{\cstdb}     {\ensuremath{0.90\pm0.32\pm0.11}}
\newcommand{\csteb}     {\ensuremath{1.14\pm0.40\pm0.14}}
\newcommand{\cstfb}     {\ensuremath{1.28\pm0.53\pm0.15}}
%
%
%
\newcommand{\Trito}       {\ensuremath{F_{(2,1)}}}      
\newcommand{\Tritz}       {\ensuremath{F_{(2,0)}}}      
\newcommand{\Trioo}       {\ensuremath{F_{(1,1)}}}      
\newcommand{\Tritt}       {\ensuremath{F_{(2,2)}}}      
\newcommand{\Tritat}      {\ensuremath{F_{(A B)}}}      
\newcommand{\Tritnt}      {\ensuremath{F_{(A \overline{B})}}}
\newcommand{\Trintat}     {\ensuremath{F_{(\overline{A} B)}}}
\newcommand{\Tritod}      {\ensuremath{56.5\pm3.9}}       
\newcommand{\Tritzd}      {\ensuremath{11.2\pm2.5}}       
\newcommand{\Triood}      {\ensuremath{\pz6.8\pm2.0}}     
\newcommand{\Trittd}      {\ensuremath{25.5\pm3.4}}       
\newcommand{\Tritofm}     {\ensuremath{45.0\,\pm\,2.5}}   
\newcommand{\Tritzfm}     {\ensuremath{16.0\,\pm\,1.7}}   
\newcommand{\Trioofm}     {\ensuremath{\pz7.3\,\pm\,1.7}} 
\newcommand{\Trittfm}     {\ensuremath{31.7\,\pm\,2.5}}   
%
%
%
%
%
%
\newcommand{\SiLaa}        {\ensuremath{0.003}}    
%
%
%
\newcommand{\nea}         {\ensuremath{6}}      
\newcommand{\neb}         {\ensuremath{4}}
\newcommand{\nec}         {\ensuremath{2}}
\newcommand{\ned}         {\ensuremath{2}}
\newcommand{\nee}         {\ensuremath{6}}
\newcommand{\nef}         {\ensuremath{4}}
%
%
\newcommand{\neaaa}       {\ensuremath{49}}      
\newcommand{\nebaa}       {\ensuremath{43}}
\newcommand{\necaa}       {\ensuremath{20}}
\newcommand{\nedaa}       {\ensuremath{16}}
\newcommand{\neeaa}       {\ensuremath{6}}
\newcommand{\nefaa}       {\ensuremath{3}}
%
%
%
\newcommand{\neab}        {\ensuremath{23}}      
\newcommand{\nebb}        {\ensuremath{23}}
\newcommand{\necb}        {\ensuremath{5}}
\newcommand{\nedb}        {\ensuremath{8}}
\newcommand{\neeb}        {\ensuremath{8}}
\newcommand{\nefb}        {\ensuremath{6}}
%
%
%
%

\begin{titlepage}
\begin{center}{\large   EUROPEAN ORGANIZATION FOR NUCLEAR RESEARCH
}\end{center}\bigskip

\begin{flushright}
       CERN-EP/2002-056   \\ 17-July-2002
\end{flushright}
\bigskip\bigskip\bigskip\bigskip\bigskip
 
\begin{center}{\huge\bf
 \boldmath Measurement of the Cross-Section for \\
           the Process $\GG\to \PP$ at \\ \vskip 3mm
           $\SQS = 183-189\,\GV$ at LEP   
 \unboldmath}
\end{center}\bigskip
\begin{center}{\LARGE The OPAL Collaboration}
\end{center}
\bigskip\bigskip
\begin{center}{\large  Abstract}
\end{center}

 The exclusive production of proton-antiproton pairs in the collisions
 of two quasi-real photons has been studied using 
 data taken at $\SQS = 183\,\GV$ and $189\,\GV$ with the OPAL detector at LEP.
 Results are presented for $\PP$ invariant masses, $W$, in the range 
 $2.15<W<3.95\,\GV$. 
 The cross-section measurements are compared with previous data and with 
 recent analytic calculations based on the quark-diquark model.
 \bigskip\bigskip\bigskip\bigskip\bigskip\bigskip

\begin{center}{\large (To be submitted to Eur. Phys. J. C.)}
\end{center}

\end{titlepage}
\begin{center}{\Large        The OPAL Collaboration
}\end{center}\bigskip
\begin{center}{
G.\thinspace Abbiendi$^{  2}$,
C.\thinspace Ainsley$^{  5}$,
P.F.\thinspace {\AA}kesson$^{  3}$,
G.\thinspace Alexander$^{ 22}$,
J.\thinspace Allison$^{ 16}$,
P.\thinspace Amaral$^{  9}$, 
G.\thinspace Anagnostou$^{  1}$,
K.J.\thinspace Anderson$^{  9}$,
S.\thinspace Arcelli$^{  2}$,
S.\thinspace Asai$^{ 23}$
% ...
}\end{center}\bigskip
\bigskip
$^{  1}$School of Physics and Astronomy, University of Birmingham,
Birmingham B15 2TT, UK
\newline
$^{  2}$Dipartimento di Fisica dell' Universit\`a di Bologna and INFN,
I-40126 Bologna, Italy
\newline
$^{  3}$Physikalisches Institut, Universit\"at Bonn,
D-53115 Bonn, Germany
\newline
$^{  4}$Department of Physics, University of California,
Riverside CA 92521, USA
\newline
$^{  5}$Cavendish Laboratory, Cambridge CB3 0HE, UK
\newline
$^{  6}$Ottawa-Carleton Institute for Physics,
Department of Physics, Carleton University,
Ottawa, Ontario K1S 5B6, Canada
\newline
$^{  8}$CERN, European Organisation for Nuclear Research,
CH-1211 Geneva 23, Switzerland
\newline
$^{  9}$Enrico Fermi Institute and Department of Physics,
University of Chicago, Chicago IL 60637, USA
\newline
$^{ 10}$Fakult\"at f\"ur Physik, Albert-Ludwigs-Universit\"at 
Freiburg, D-79104 Freiburg, Germany
\newline
$^{ 11}$Physikalisches Institut, Universit\"at
Heidelberg, D-69120 Heidelberg, Germany
\newline
$^{ 12}$Indiana University, Department of Physics,
Swain Hall West 117, Bloomington IN 47405, USA
\newline
$^{ 13}$Queen Mary and Westfield College, University of London,
London E1 4NS, UK
\newline
$^{ 14}$Technische Hochschule Aachen, III Physikalisches Institut,
Sommerfeldstrasse 26-28, D-52056 Aachen, Germany
\newline
$^{ 15}$University College London, London WC1E 6BT, UK
\newline
$^{ 16}$Department of Physics, Schuster Laboratory, The University,
Manchester M13 9PL, UK
\newline
$^{ 17}$Department of Physics, University of Maryland,
College Park, MD 20742, USA
\newline
$^{ 18}$Laboratoire de Physique Nucl\'eaire, Universit\'e de Montr\'eal,
Montr\'eal, Quebec H3C 3J7, Canada
\newline
$^{ 19}$University of Oregon, Department of Physics, Eugene
OR 97403, USA
\newline
$^{ 20}$CLRC Rutherford Appleton Laboratory, Chilton,
Didcot, Oxfordshire OX11 0QX, UK
\newline
% ...
$^{ 22}$Department of Physics and Astronomy, Tel Aviv University,
Tel Aviv 69978, Israel
\newline
$^{ 23}$International Centre for Elementary Particle Physics and
Department of Physics, University of Tokyo, Tokyo 113-0033, and
Kobe University, Kobe 657-8501, Japan
\newline
$^{ 24}$Particle Physics Department, Weizmann Institute of Science,
Rehovot 76100, Israel
\newline
$^{ 25}$Universit\"at Hamburg/DESY, Institut f\"ur Experimentalphysik, 
Notkestrasse 85, D-22607 Hamburg, Germany
\newline
$^{ 26}$University of Victoria, Department of Physics, P O Box 3055,
Victoria BC V8W 3P6, Canada
\newline
$^{ 27}$University of British Columbia, Department of Physics,
Vancouver BC V6T 1Z1, Canada
\newline
$^{ 28}$University of Alberta,  Department of Physics,
Edmonton AB T6G 2J1, Canada
\newline
$^{ 29}$Research Institute for Particle and Nuclear Physics,
H-1525 Budapest, P O  Box 49, Hungary
\newline
$^{ 30}$Institute of Nuclear Research,
H-4001 Debrecen, P O  Box 51, Hungary
\newline
$^{ 31}$Ludwig-Maximilians-Universit\"at M\"unchen,
Sektion Physik, Am Coulombwall 1, D-85748 Garching, Germany
\newline
$^{ 32}$Max-Planck-Institut f\"ur Physik, F\"ohringer Ring 6,
D-80805 M\"unchen, Germany
\newline
$^{ 33}$Yale University, Department of Physics, New Haven, 
CT 06520, USA
\newline
\bigskip\newline
% ...
$^{  c}$ and Institute of Nuclear Research, Debrecen, Hungary
\newline
% ...
$^{  e}$ and Department of Experimental Physics, Lajos Kossuth University,
 Debrecen, Hungary
\newline
% ...

\section{Introduction}
\label{sec:introduction}
The exclusive production of proton-antiproton (\PP) pairs in the collision of
two quasi-real photons can be used to test predictions of
QCD. At LEP the photons are emitted by the beam electrons\footnote{In
this paper positrons are also referred to as electrons.}
and
the $\PP$ pairs are produced in the process $\EE\to\EE\gamma\gamma\to\EE\PP$.

The application of QCD to exclusive photon-photon reactions
is based on the work of Brodsky and Lepage~\cite{Lepage:1980fj}. 
According to their formalism the process is factorized into a 
non-perturbative part, which is the hadronic wave function of the final 
state, and a perturbative part.
Calculations based on this ansatz~\cite{Farrar:1985gv,Millers:1986ca}
use a specific model of the proton's three-quark wave function by 
Chernyak and Zhitnitsky~\cite{Chernyak:1984bm}.
This calculation yields cross-sections about one order of magnitude smaller 
than the existing experimental results~\cite{Althoff:1983pf,Bartel:1986sy,Aihara:1987ha,Albrecht:1989hz,Artuso:1994xk,Hamasaki:1997cy}, 
for $\PP$ centre-of-mass energies $W$ greater than $2.5\,\GV$.

To model non-perturbative effects, the introduction
of quark-diquark systems has been proposed~\cite{Ansel:1987vk}. 
Within this model, baryons are viewed as a combination of a quark and 
a diquark rather than a three-quark system.
The composite nature of the diquark is taken into account by form factors.

Recent studies~\cite{berger:1997} have extended the systematic 
investigation of hard exclusive reactions within the quark-diquark model 
to photon-photon
processes~\cite{Anselmino:1989gu,Kroll:1991a,Kroll:1993zx,Kroll:1996pv}.
In these studies the cross-sections have been calculated down to $W$ 
values of $2.2\,\GV$ below which the quark-diquark model is no longer expected
to be valid. Most of the experimental data, however, have 
been taken at such low energies.

The calculations of the integrated cross-section for the process $\GG\to\PP$
in the angular range $\costs < 0.6$ (where $\theta^{*}$ is the angle between 
the proton's momentum and the electron beam direction in the  
$\PP$ centre-of-mass system) and for $W > 2.5\,\GV$ are 
in good agreement with experimental 
results~\cite{Artuso:1994xk,Hamasaki:1997cy}, whereas the pure quark model 
predicts much smaller cross-sections~\cite{Farrar:1985gv,Millers:1986ca}.  

In this paper, we present a measurement of the cross-section 
for the exclusive process $\EE\to\EE \PP$ in the range 
$2.15<W<3.95\,\GV$, using data taken with the OPAL detector 
at $\SQS = 183\,\GV$ and $189\,\GV$ at LEP. 
The integrated luminosities for the 
two energies are $62.8\,\pb^{-1}$ and $186.2\,\pb^{-1}$.


\section{The OPAL detector} 
\label{sec:OPAL}

The OPAL detector and trigger system are described in detail
elsewhere~\cite{Ahmet:1991eg}. 
We briefly describe only those features particularly relevant to
this analysis.
The tracking system for charged particles is inside a 
solenoid that provides a uniform axial magnetic field of $0.435\,{\rm T}$. 
The system consists of a silicon micro-vertex detector, a 
high-resolution vertex drift chamber, a large-volume jet chamber and 
surrounding $z$-chambers.
The micro-vertex detector surrounds the beam pipe covering the angular 
range of $\cost<0.9$ and provides tracking information in the 
$r$-$\phi$ and $z$ directions\footnote{In the OPAL right-handed coordinate 
system the $z$-axis points along the ${\rm e}^-$ beam direction, and the 
$x$-axis points towards the centre of the LEP ring.  The polar angle 
$\theta$ is defined with respect to the $z$-axis, and the azimuthal angle 
$\phi$ with respect to the $x$-axis.}. 
The jet chamber records the momentum and energy loss of charged particles 
over $98\%$ of the solid angle.
In the range of $\cost < 0.73$, up to $159$ points are measured along 
each track.
The energy loss, $\DEDX$, of a charged particle in the chamber gas is measured 
from the integrated charges of each hit at both ends of each signal wire 
with a resolution of about $3.5\%$ for isolated tracks with the maximum of 
159 points.
The $z$-chambers are used to improve the track measurement in the $z$ 
direction.

The barrel time-of-flight (TOF) scintillation counters
are located immediately outside the solenoid at a mean radius of
$2.36\,\m$, covering the polar angle range $\cost < 0.82$.
The outer parts of the detector, in the barrel and endcaps, consist of 
lead-glass electromagnetic calorimetry (ECAL) followed by an instrumented 
iron yoke for hadron calorimetry and four layers of external muon chambers.
Forward electromagnetic calorimeters complete the acceptance
for electromagnetically interacting particles down to polar angles of
about $24\,\mrad$.

The trigger signatures required for this analysis are based on a combination
of time-of-flight and track trigger information.


\section{Kinematics} 
\label{sec-kine}

The production of proton-antiproton pairs in photon-photon
interactions proceeds via the process
\begin{equation}
\label{eq:kin}
\mathrm{e}^{+}(p_{1}) + \mathrm{e}^{-}(p_{2})\to
\mathrm{e}^+(p_{1}') + \mathrm{e}^-(p_{2}')+\gamma(q_1) + \gamma(q_2)\to 
\mathrm{e}^+(p_{1}') + \mathrm{e}^-(p_{2}') + \mathrm{p}(q_{1}') 
+ \mathrm{\bar{p}}(q_{2}')
\end{equation}
where $q_{i},p_{i}$ denote the four-momenta ($i=1,2$).
Each of the two incoming electrons emits a photon and
the final state produced by 
the two colliding photons consists of one proton (p) and one
antiproton ($\overline{\mbox{p}}$). 
The four-momentum squared of the two photons is ($i=1,2$)
\begin{equation}
 \label{eq:q2}
  q_i^{2}=-Q_i^{2}=(p_i - p_{i}')^{2}.
\end{equation}
Since the electrons are scattered at small angles and they
remain undetected, the four-momenta squared of each of the
two photons are small, i.e.~the photons are quasi-real.
In this case the transverse\footnote{
In this paper transverse momenta are always defined with respect to the $z$ 
axis.} 
component of the momentum sum of the
proton and the antiproton in the laboratory system is expected to 
be small whereas the longitudinal component of the momentum sum
can be large.  
The photon-photon centre-of-mass system is generally boosted along 
the beam axis. 
The larger the boost, the closer the produced (anti) protons are to 
the beam direction.  
This feature, combined with the typically low mass of the 
final state $\PP$, and the low efficiency for tracking at small angles, 
leads to significant acceptance losses for these types of events.


\section{Monte Carlo generators} 
\label {sec:simulation}
The $\EE\to\EE\PP$ events are simulated with the PC Monte Carlo 
generator which has been developed to study 
exclusive photon-photon processes~\cite{linde}.
The PC Monte Carlo generator has been expanded for 
use in this analysis to simulate the kinematics of 
exclusive baryon-antibaryon final states, $\EE\to\EE\PP$, 
$\EE\to\EE\LL$, and $\EE\to\EE\XIXI$.
A total of $20\,000$ $\EE\to\EE\PP$ Monte Carlo events have been generated 
as a control sample for the event selection described in the next section. 
Similarly, for the trigger and detection efficiency determination, a total 
of $360\,000$ $\EE\to\EE\PP$ Monte Carlo events have been generated 
in the range of $2.15<W<3.95\,\GV$.   
The background coming from $\EE\to\EE\PiPi$ is generated with 
the PC Monte Carlo program, as is
the feed-down background from proton-antiproton pairs coming from the reaction
$\EE\to\EE\LL\to\PP\PiPi$ where the pions are not detected ($200\,000$ events).

The leptonic photon-photon background processes $\EE\to\EE\MuMu$, 
$\EE\to\EE\EE $ and $\EE\to\EE\TT$
are simulated with the Vermaseren generator~\cite{Vermaseren:1983cz}. 
The KORALZ generator~\cite{Jadach:1994yv} is used to simulate the 
background processes $\EE\to\MuMu$, and $\EE\to\TT$.
The $\EE\to\EE$ background process is simulated with the
BHWIDE~\cite{Jadach:1997nk} generator.
Table~\ref{tab:mcbg} lists all the generated background Monte Carlo samples. 

Monte Carlo events are generated at $\SQS = 189\,\GV$ only, since
the change in acceptance between $\SQS = 183$ and $189$~GeV is small
compared to the statistical uncertainty of the measurement.
They have been processed through a full simulation of the OPAL 
detector~\cite{Allison:1992bf} and have been analysed using 
the same reconstruction algorithms that are used for the data.


\section{Event selection} 
\label {sec:selection} 

\begin{figure}[p]
 \centering
  \resizebox{0.6\textwidth}{!}{%
   \includegraphics{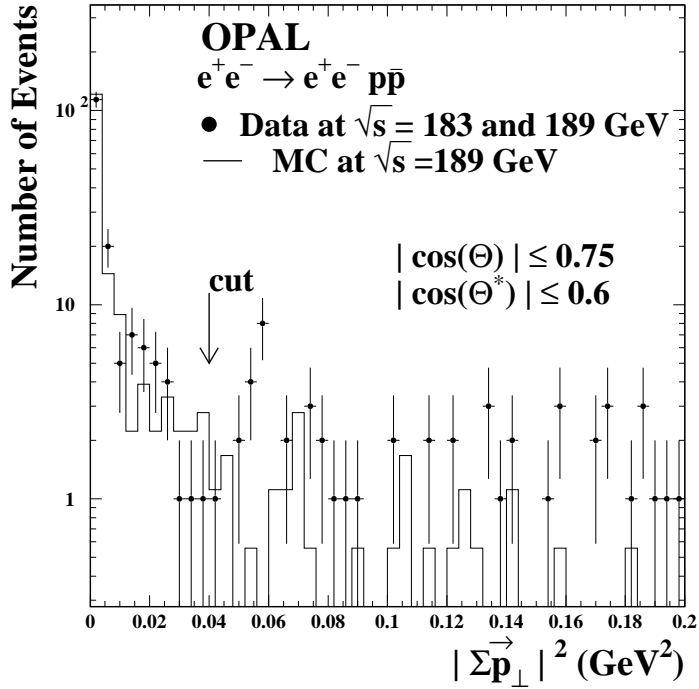}}
    \caption{Distribution of the transverse momentum balance, $\spts$, 
         for the $\EE\to\EE\PP$ events in data (black points with error 
         bars) compared to the Monte Carlo simulation (histogram).
         The distributions are obtained after applying all 
         cuts except the $\spts$ cut.
         The arrow indicates the cut value. The error bars
         are statistical only. The Monte Carlo distribution   
         is normalized to the number of selected
         data events.}\label{fig:spt2} 
\end{figure}

The $\EE\to\EE\PP$ events are selected by the following set of cuts:
\begin{enumerate}
\item
     The sum of the energies measured in the barrel and endcap sections
     of the electromagnetic calorimeter must be less than half the beam
     energy.  
\item
     Exactly two oppositely charged tracks are required with each track
     having at least 20 hits in the central jet chamber to ensure
     a reliable determination of the specific energy loss ${\rm d}E/{\rm d}x$.
     The point of closest approach to the interaction point
     must be less than 1~cm in the $r\phi$ plane 
     and less than 50~cm in the $z$ direction.
\item
     For each track the polar angle must be in the range $\cost < 0.75$ 
     and the transverse momentum
     $\PT$ must be larger than $0.4\,\GV$.
     These cuts ensure a high 
     trigger efficiency and good particle identification.
\item 
     The invariant mass $W$ of the $\PP$ final state
     must be in the range $2.15<W<3.95$~GeV. 
     The invariant mass is determined from the measured 
     momenta of the two tracks using the proton mass.
\item
     The events are boosted into the rest system of the
     measured $\PP$ final state.
     The scattering angle of the tracks in this
     system has to satisfy $\costs < 0.6$.
\item 
     All events must fulfil the trigger
     conditions defined in Section~\ref{sec:trigger}.
\item
     The large background from other exclusive processes, 
     mainly the production of e$^+$e$^-$, $\mu^+\mu^-$, and $\pi^+\pi^-$
     pairs, is reduced by particle identification 
     using the specific energy loss $\dedx$ in the jet
     chamber and the energy in the electromagnetic calorimeter. The 
     $\dedx$ probabilities of the tracks must be consistent with the 
     p and $\overline{\mbox{p}}$ hypothesis. 
     \begin{itemize}
     \item[-]
      Events where the ratio $E/p$ for each track lies in the range 
      $0.4 < E/p < 1.8$~\footnote{$E$ here is the energy of the ECAL cluster 
      associated with the track with momentum~$p$.} are regarded as possible
      $\EE\to\EE\EE$ candidates. These events are rejected if the 
      $\dedx$ probabilities of the two tracks are consistent 
      with the electron hypothesis.
      \item[-]
      Events where the ratio $E/p$ for each track is less than $0.8$, 
      as expected for a minimum ionizing particle, are regarded as possible
      background from $\EE\to\EE\mu^+\mu^-$ events. This background is 
      reduced by rejecting events where the $\dedx$ probability for
      both tracks is consistent with the muon hypothesis. 
      This cut is also effective in reducing the $\pi^{+}\pi^{-}$
      background.      
     \item[-]
      The $\dedx$ probability for the proton hypothesis has to be greater than 
      $0.1\%$ for each track and it has to be larger than the probabilities 
      for the pion and kaon hypotheses. 
      The product of the $\dedx$ probabilities for both 
      tracks to be (anti) protons has to be larger than the product of the 
      $\dedx$ probabilities for both tracks to be electrons.
     \end{itemize}
\item
     Cosmic ray background is eliminated by applying a 
     muon veto~\cite{Akers:1995vh}.
\item
     Exclusive two-particle final states are selected by
     requiring the transverse component of the momentum 
     sum squared of the two tracks, $\spts$, to be smaller than
     $0.04\,\GV^2$. By restricting the maximum value of $Q_i^2$, 
     this cut also ensures that the interacting photons
     are quasi-real. Therefore no further cut rejecting events with
     scattered electrons in the detector needs to be applied.
     Fig.~\ref{fig:spt2} shows the $\spts$ distribution obtained after 
     applying all cuts except the cut on $\spts$.
\end{enumerate}

\begin{figure}[htbp]
\centering
\resizebox{0.6\textwidth}{!}{%
  \includegraphics{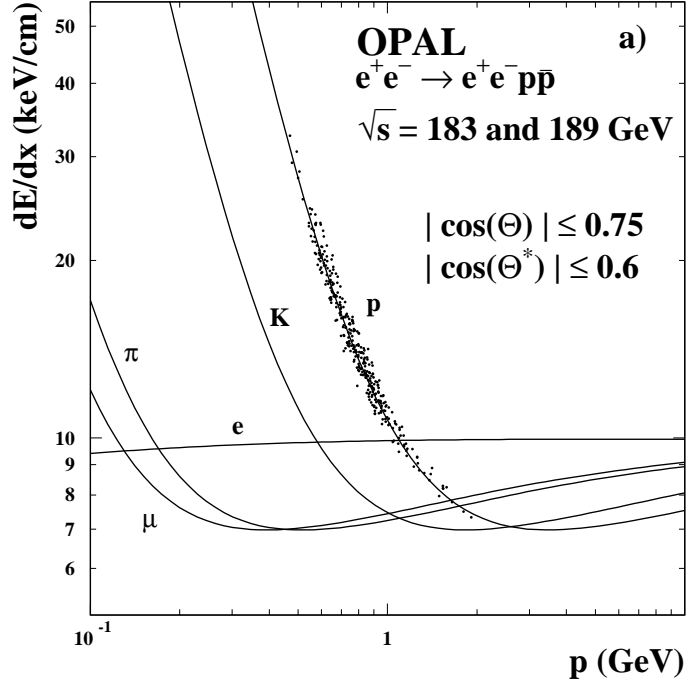}} 
  \resizebox{0.6\textwidth}{!}{%
   \includegraphics{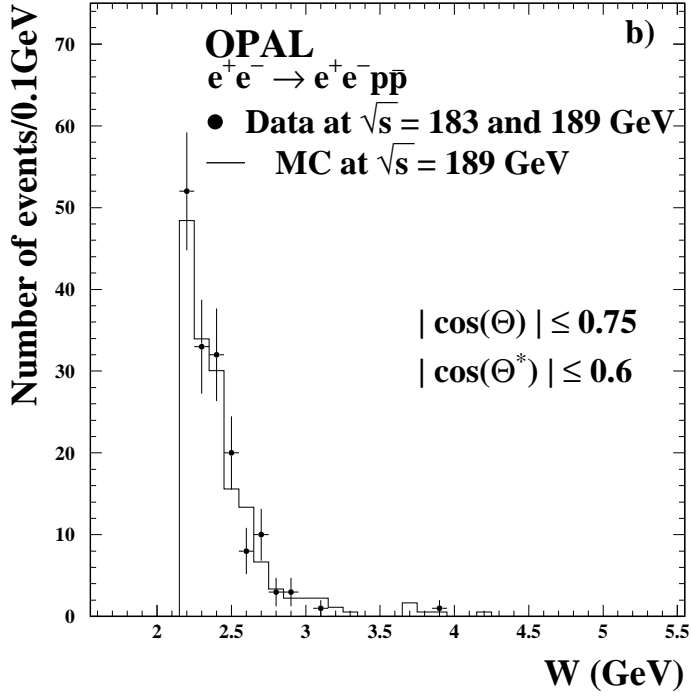}}
     \caption{ 
         a) Distribution of the specific energy loss,
         $\dedx$, versus the particle momentum $p$
         for the proton and antiproton tracks selected 
         by applying the final selection.
         The curves indicate the expected mean values for
         different particle species.
         b) Invariant mass distribution for the 163 selected 
         $\EE\to\EE\PP$ events.
         The dots with error bars are the data,
         the histogram denotes Monte Carlo events. Errors bars are
         statistical only. The Monte Carlo distribution  
         is normalized to the number of selected data events.
         }
\label{fig:sel6} 
\end{figure}

After all cuts $163$ data events are selected, 35 events 
at $\SQS=183$~GeV and 128 events at $\SQS=189$~GeV.
The distribution of measured 
$\dedx$ values versus the particle momentum for the selected data events 
is shown in Fig.~\ref{fig:sel6}a.
Background from events containing particles other than (anti-)protons
is negligible due to the good rejection power of the $\dedx$ cuts.
Since no event remains
after applying the event selection to the background Monte Carlo
samples, the $95\%$ confidence level upper limits 
for the number of background events expected to contribute  
to the selected data sample are given in Table~\ref{tab:mcbg}. 

Since the $\PP$ final state is fully reconstructed, the
experimental resolution for $W$ (determined with Monte Carlo simulation)
is better than $1\%$. The experimental resolution for $\costs$ is about 
$0.014$.
Fig.~\ref{fig:sel6}b shows the $W$ distribution for data and Monte 
Carlo signal events after the final selection. The Monte Carlo
distributions agree well with the data. 


\section{Trigger and detection efficiencies}   
\label {sec:trigger}

The $\EE\to\EE\PP$ events contain only two tracks  
with momenta in the range 0.4 GeV to 2~GeV. 
Special triggers 
are required to select such low multiplicity events with
only low momentum particles and
the efficiencies of these triggers must be well known.
The $\EE\to\EE\PP$ events are mainly triggered by a  
combination of triggers using  
hits in the time-of-flight counters and tracks.
The track trigger takes data on the $z$ coordinate of hits 
from the vertex drift chamber, and from three
groups of 12 wires in the jet chamber.
The selected events must satisfy at least one of the following trigger 
conditions:
\begin{itemize}
\item[A]
     Two tracks in the barrel region from the track-trigger ({\rm TT}). 
     This corresponds to an angular acceptance of approximately 
     $\cost < 0.75$.
\item[B]
     A coincidence of at least one barrel track from the 
     track trigger and a $\TP$ coincidence of a track from the
     track trigger with hits from the time-of-flight detector ({\rm TOF}).
     The barrel track and the track forming the $\TP$
     coincidence are not necessarily identical.
     The angular acceptance of the track trigger is  
     approximately $\cost < 0.75$, whereas the acceptance of
     the $\TP$ coincidence is $\cost < 0.82$.
\end{itemize}
Condition A is highly efficient for events within its
geometrical acceptance but it triggers on two tracks. 
Condition B is used to trigger on a single track and 
to measure the trigger efficiency in combination with condition A.

It was checked that the trigger efficiency does not depend
on $\phi$. Under this condition and assuming the efficiency of each 
trigger component for each track to be independent, the combined event 
trigger can be written as:
\begin{equation}
 \ETRIG= \varepsilon_{\rm TT}^2+2\varepsilon_{\rm TT-TOF}
         \varepsilon_{\rm TT}(1-\varepsilon_{\rm TT}).
\label{eq:trig}
\end{equation}
Here $\varepsilon_{\rm TT}$ is the efficiency for one track to be 
triggered by the track trigger and $\varepsilon_{\rm TT-TOF}$ 
is the efficiency for one track to be triggered by the TT-TOF
coincidence.
The first term in (\ref{eq:trig}) gives the efficiency 
for both tracks to be triggered by the track trigger (condition A) and
the second term gives the efficiency for the events
not triggered by condition A to be triggered by condition B 
for either of the two tracks.

Data events are used to calculate the trigger efficiency.
The trigger efficiency is determined by considering events in which
one track detected in one half of the $r\phi$ plane (e.g. 
$0<\phi<180$~degrees) satisfies the required track trigger and 
time-of-flight $\TP$ matching hit while the other track in the other half 
plane (e.g. $180<\phi<360$ degrees) is used to measure the efficiency of the 
track and time-of-flight triggers. 

The efficiencies measured from the $163$ selected events
are $\varepsilon_{\rm TT} = (93.7\pm1.7)\%$
and $\varepsilon_{\rm TT-TOF} = (58.5\pm2.3)\%$, where the
uncertainties are statistical only. This yields an overall event 
trigger efficiency of $\varepsilon_{\rm TRIG} = (94.7\pm1.5)\%$, 
determined from data only.

To study the efficiency with a large event sample, events 
from photon-photon processes with two tracks in the final state 
such as $\EE$ and $\MuMu$ are used together with
the $\PP$ events to determine the track trigger efficiency as 
a function of $\PT$. 
This efficiency exceeds $90\%$ for $\PT>0.4\,\GV$ and the observed
dependence of the track trigger efficiency on $\PT$ is found to be consistent 
between electrons and muons.

In a second step, $\EE\to\EE\PP$ Monte Carlo events in the 
range $2.15<W<3.95\,\GV$ are used to obtain the trigger 
efficiencies as a function of $W$, $\varepsilon_{\rm TRIG}(W)$. 
The Monte Carlo events are reweighted here according to the 
trigger efficiency which has been determined as a function of the
transverse momentum $\PT$ using~(\ref{eq:trig}). These reweighted
Monte Carlo events have been used only to determine the trigger 
efficiency as function of $W$.  
Finally the values obtained for $\varepsilon_{\rm TRIG}(W)$
are normalized to give the overall trigger efficiency
$\varepsilon_{\rm TRIG}=94.7\%$ above the threshold 
$\PT>0.4$~GeV as already determined from data only.
The region $W<2.15\,\GV$ is excluded from the analysis because the 
trigger efficiency drops rapidly below $70\%$. 

The detection efficiency is determined by comparing the number of Monte 
Carlo events passing all cuts with the total number of Monte Carlo events 
generated within a polar angle $\costsgen < 0.6$ in the $\PP$ 
centre-of-mass system:
\begin{equation}
\EDET =\frac{ {\rm d}\sigma/{\rm d}W(|\cos\theta^{*}|)}{{\rm d}
\sigma/{\rm d}W(|\cos\theta^{*}_{\rm GEN}|)},
\label{eq:deteff}
\end{equation}
where $\costs$ refers to the reconstructed
polar angle in the $\PP$ centre-of-mass system.
The detection efficiency is about $2\%$ at high $W$ and about 
$4\%$ at low $W$.
To be able to compare the measured cross-section with any given model
in bins of $\costs$ and $W$, the detection efficiency has been determined 
from the signal Monte Carlo in bins of $W$ and $\costs$.

\section{Cross-section measurements}
\label{sec:cross section}
The differential cross-section for the process $\EE\to\EE\PP$ is given by
\begin{equation}
  \frac{{\rm d}^2\SI(\EE\to\EE\PP)}{{\rm d}W\,{\rm d}\costs} = 
  \frac{{\NE(W,\costs)}}{{\cal{L}}_{\EE}\ETRIG\,\EDET\,(W,\costs)\,\Delta W\,\Delta\costs}
 \label{eq:diffcross}
\end{equation}
where \NE\ is the number of events selected in each $(W,\costs)$
bin, $\ETRIG$ is the trigger efficiency, $\EDET$ is the detection 
efficiency, $\cal{L}_{\EE}$ is the measured integrated luminosity, and 
$\Delta W$ and $\Delta\costs$ are the bin widths in $W$ and in $\costs$.

The total cross-section $\SI(\GG\to\PP)$ for a given value of 
$\SQS$ is obtained from the differential cross-section
${\rm d}\SI(\EE\to\EE\PP)/{\rm d}W$ using the luminosity
function ${\rm d}\lumi_{\GG}/{\rm d}W$~\cite{Low:1960wv}:
\begin{equation}
 \SI(\GG\to\PP)= 
 \frac{{\rm d}\SI(\EE\to\EE\PP)}{{\rm d}W}\left/ 
 \frac{{\rm d}\lumi_{\GG}}{{\rm d}W}\right. .
 \label{eq:cross}
\end{equation}
The luminosity function ${\rm d}\lumi_{\GG}/{\rm d}W$ 
is calculated by the {\sc Galuga} program~\cite{Schuler:1996gt}.
The resulting differential cross-sections for the process $\GG\to\PP$ 
in bins of $W$ and $\costs$ are then summed over 
$\costs$ to obtain the total cross-section as a function of $W$
for $\costs<0.6$.

\section{Systematic uncertainties}
\label{sec:systematics}
The following sources of systematic uncertainties have been taken
into account (Table~\ref{tab:sys}):
\begin{description}
\item[Luminosity function:]
     The accuracy of the photon-photon luminosity function used 
     in (\ref{eq:cross}) has been estimated by taking into account 
     different models such as the $\rho$-pole model~\cite{Buijs:1994wi}, 
     the Equivalent Photon Approximation (EPA)~\cite{art:fermi}, 
     the Generalized Vector Dominance Model (GVMD)~\cite{Schuler:1996gt},
     the luminosity functions given in~\cite{Field:1980pf}, and 
     in~\cite{Budnev:1974de,Brodsky:1971ud}. 
     Each of the resulting numbers has then been compared with the VDM 
     form-factor model~\cite{Schuler:1996gt} which is used for
     the final result.
     We take the largest deviation resulting from these comparisons as 
     the uncertainty in the luminosity functions which is about $\pm5\%$. 
\item [Trigger efficiency:]
      The trigger efficiency model of (\ref{eq:trig}) has been
      tested by making a comparison between the measured relative 
      frequencies of the four different sub-combinations, 
      ($N_{\rm TT},N_{\rm TT-TOF}$) = (1,1), (2,0), (2,1), (2,2),
      which can trigger the events, and the predicted values, by using
      fitted efficiencies for TT and TT-TOF (Section~\ref{sec:trigger}). 
      Here $N_{\rm TT}$ denotes the number of tracks triggered by 
      the {\rm TT} and $N_{\rm TT-TOF}$ the number of tracks 
      triggered by the {\rm TT-TOF} trigger.
      Table~\ref{tab:fractions} gives the measured and predicted fractions 
      for events with  $W > 2.15\,\GV$. 
      The fit of the two efficiencies $\varepsilon_{\rm TT}$ and 
      $\varepsilon_{\rm TT-TOF}$ yields a $\chi^{2}$ of $9.6$ over $2$ 
      degrees of freedom.
      Although the fit is poor the overall efficiency result is 
      consistent with the other determination. 
      An additional systematic uncertainty of $5.0\%$ is 
      therefore assigned to the trigger efficiency.  
      This value is obtained by increasing the 
      uncertainties to obtain a normalized $\chi^{2}$ of one.  
\item[Monte Carlo statistics:]
      The statistical uncertainty on the detection
      efficiency due to the number of simulated Monte Carlo events
      varies from $4.5\%$ at low $W$ to $6\%$ at high $W$.
\item[\bf \boldmath$ {\rm d}E/{\rm d}x$ cuts:\unboldmath] 
     The systematic uncertainties due to the \DEDX\ cuts are 
     determined by recalculating the Monte Carlo detection efficiency 
     after varying the $\DEDX$ values:
     \begin{itemize}
     \item[-] The measured values are shifted by $\pm1\sigma$,
      where $\sigma$ is the theoretical uncertainty of the measurement.
     \item[-] The measured values are
      smeared with a Gaussian distribution of width $0.1\sigma$,
      which is the typical systematic uncertainty of each 
      individual \DEDX\ measurement, as found for pions 
      from $\K$ decays~\cite{Bose}.
     \end{itemize}  
     The modified $\DEDX$ values are transformed into weights and the
     new detection efficiency is calculated by applying the 
     event selection on the modified Monte Carlo events.
     The systematic uncertainty assigned to each $(W,\costs)$ bin is the 
     quadratic sum of the full deviation of the detection efficiency  
     with smearing and the average absolute deviation value of the 
     $\pm1\sigma$ shifted values with respect to the original values.
     
     The systematic uncertainties due to the variation of the $\dedx$ cuts 
     are larger at high values of $W$. They vary between $0.1\%$
     for $W< 2.6\,\GV$ and up to $5\%$ for $W > 2.6\,\GV$.
  
\item[Residual background:] 
     Residual background can come from 
     non-exclusive production of $\PP$ pairs in processes like 
     $\EE\to\EE\PP\pi\pi$ if the pions are not detected.
     The $\spts$ cut eliminates most of this background.
     The data events are almost coplanar, i.e. 
     no data event has an acoplanarity of more than $0.262$~rad.
     To estimate the contribution of this background,
     the shape of the acoplanarity distribution 
     for $\PP$ pairs in Monte Carlo $\EE\to\EE\LL$ and 
     $\EE\to\EE\PP$ events has been fitted to the data. 
     This yields an upper limit of 10 events for
     this background contribution.
     An additional systematic uncertainty 
     of $6\%$ is therefore taken into account. 
\end{description}
     Additional uncertainties due 
     to the measured integrated $\EE$ luminosity, the 
     track reconstruction efficiency and the momentum resolution 
     for the protons are negligible. 
     The total systematic uncertainty is obtained by adding 
     all systematic uncertainties in quadrature. 


\section{Results and Discussion}
\label{sec:results}

\begin{figure}[htbp]
 \centering
  \resizebox{0.48\textwidth}{!}{%
  \includegraphics{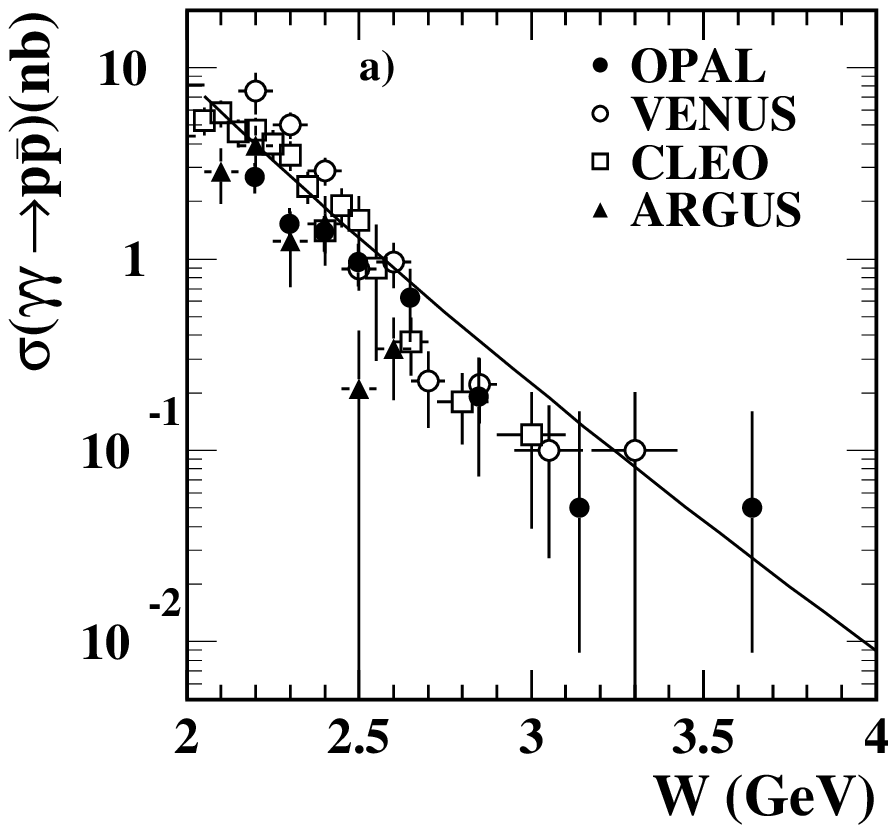}}
  \resizebox{0.48\textwidth}{!}{%
  \includegraphics{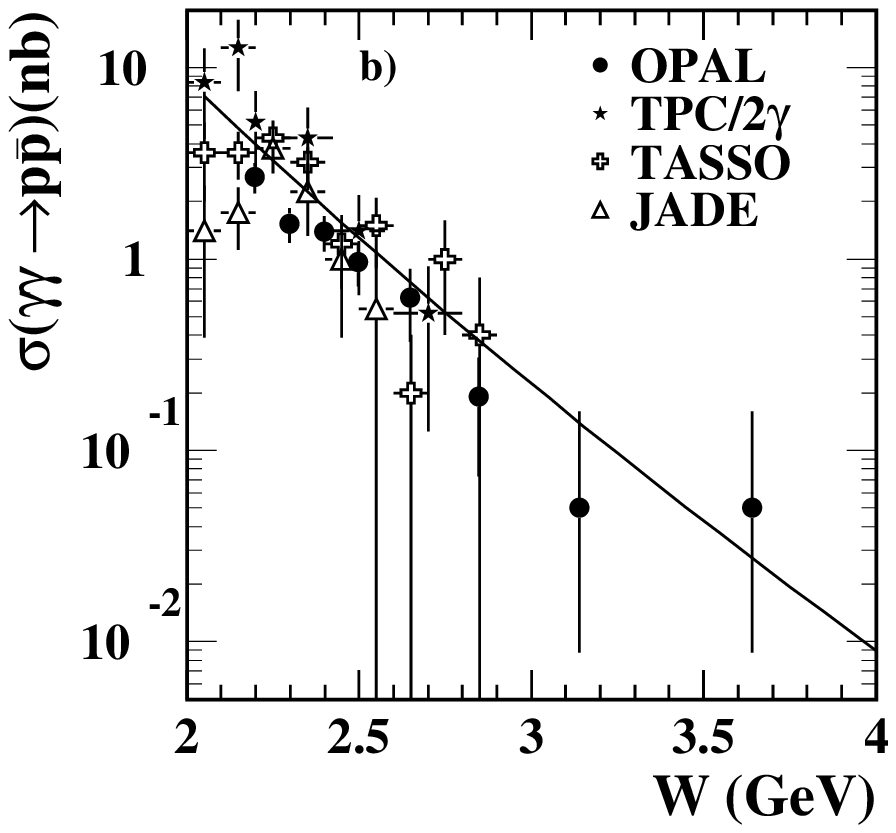}}
  \resizebox{0.48\textwidth}{!}{%
  \includegraphics{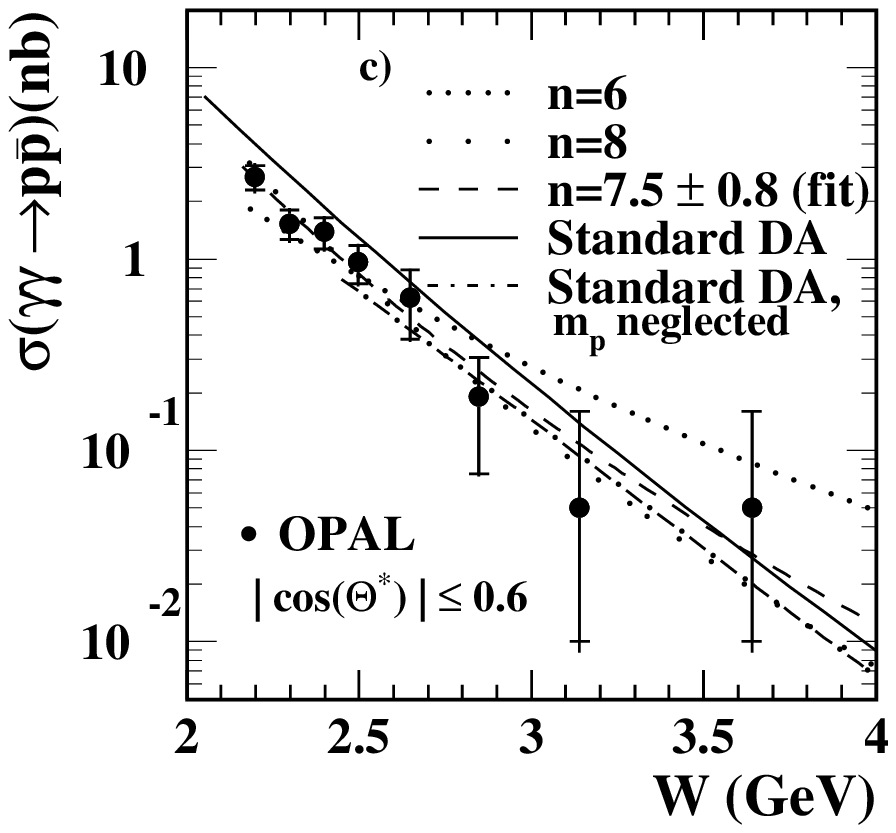}}
    \caption{Cross-sections $\sigma(\GG\to\PP)$ as a function of $W$.
        The data and the theoretical predictions cover a range of 
        $\costs < 0.6$. 
        The data points are plotted at the value of $\langle W\rangle$.
        a,b) The data are compared to other experimental 
     results~\cite{Aihara:1987ha,Albrecht:1989hz,Artuso:1994xk,Hamasaki:1997cy}
        and to the quark-diquark model prediction~\cite{berger:1997}.
        The error bars include statistical and systematic uncertainties, 
        except for TASSO~\cite{Althoff:1983pf} where the uncertainties are
        statistical only. 
        c) The data are compared to the quark-diquark model 
        of~\cite{Ansel:1987vk} (dash-dotted line), and 
        of~\cite{berger:1997} (solid line), using the standard distribution 
        amplitude (DA) with and without neglecting the mass $m_{\rm p}$
        of the proton, 
        and with the predictions of the power law with  fixed 
        and with fitted exponent $n$. 
        The inner error bars are the statistical 
        uncertainties and the outer error bars are the total uncertainties.}
    \label{fig:w}
\end{figure}

The measured cross-sections in bins of $W$ are given in 
Table~\ref{tab:crossmeas}.
The average $\langle W\rangle$ in each bin has been
determined by applying the procedure described in~\cite{Lafferty:1995}. 
The measured cross-sections $\SI(\GG\to\PP)$ 
for $2.15<W<3.95\,\GV$ and for $\costs < 0.6$ 
are compared with the results obtained by 
ARGUS~\cite{Albrecht:1989hz}, CLEO~\cite{Artuso:1994xk} and 
VENUS~\cite{Hamasaki:1997cy} in Fig.~\ref{fig:w}a
and to the results obtained by
TASSO~\cite{Althoff:1983pf}, 
JADE~\cite{Bartel:1986sy} and TPC/$2\gamma$~\cite{Aihara:1987ha}
in Fig.~\ref{fig:w}b.
The quark-diquark model predictions~\cite{berger:1997} are also shown.
Reasonable agreement is found between this measurement 
and the results obtained by other experiments for $W>2.3\,\GV$.
At lower $W$ our measurements agree with the measurements by 
JADE~\cite{Bartel:1986sy} and ARGUS~\cite{Albrecht:1989hz},
but lie below the results
obtained by CLEO~\cite{Artuso:1994xk}, and VENUS~\cite{Hamasaki:1997cy}.
The cross-section measurements reported here extend towards higher 
values of $W$ than previous results.

Fig.~\ref{fig:w}a-b show the measured $\GG\to\PP$ 
cross-section as a function of $W$ together with some  
predictions based on the quark-diquark 
model~\cite{Ansel:1987vk,berger:1997,Kroll:1993zx}.  
There is good agreement between our results and the older
quark-diquark model predictions~\cite{Ansel:1987vk,Kroll:1993zx}.
The most recent calculations~\cite{berger:1997} lie above the data, 
but within the estimated theoretical uncertainties the predictions are 
in agreement with the measurement.

An important consequence of the pure quark hard scattering 
picture is the power law which follows from the dimensional counting 
rules~\cite{Brodsky:1973kr,Matveev:1973ra}.
The dimensional counting rules state that an exclusive cross-section 
at fixed angle has an energy dependence connected with the number of 
hadronic constituents participating in the process under investigation. 
We expect that for asymptotically large $W$ and fixed 
$\costs$
\begin{equation}
  \frac{{\rm d}\SI{(\GG\to\PP)}}{{\rm d}t} \sim W^{2(2-n)}
  \label{eq:powerlaw}
\end{equation}
where $n=8$ is the number of elementary fields
and $t = -W^2/2(1-\costs)$. The introduction
of diquarks modifies the power law by decreasing $n$ to $n=6$.
This power law is compared
to the data in Fig.~\ref{fig:w}c with 
$\SI(\GG\to\PP) \sim W^{-2(n-3)}$ using three 
values of the exponent $n$: fixed values $n=8$, $n=6$, 
and the fitted value 
$n=7.5\pm0.8$ obtained by taking into account statistical uncertainties only. 
More data covering a wider range of $W$ would be required to determine the 
exponent $n$ more precisely. 

\begin{center}
 \begin{figure}[htbp]
   \centering
   \resizebox{0.48\textwidth}{!}{%
   \includegraphics{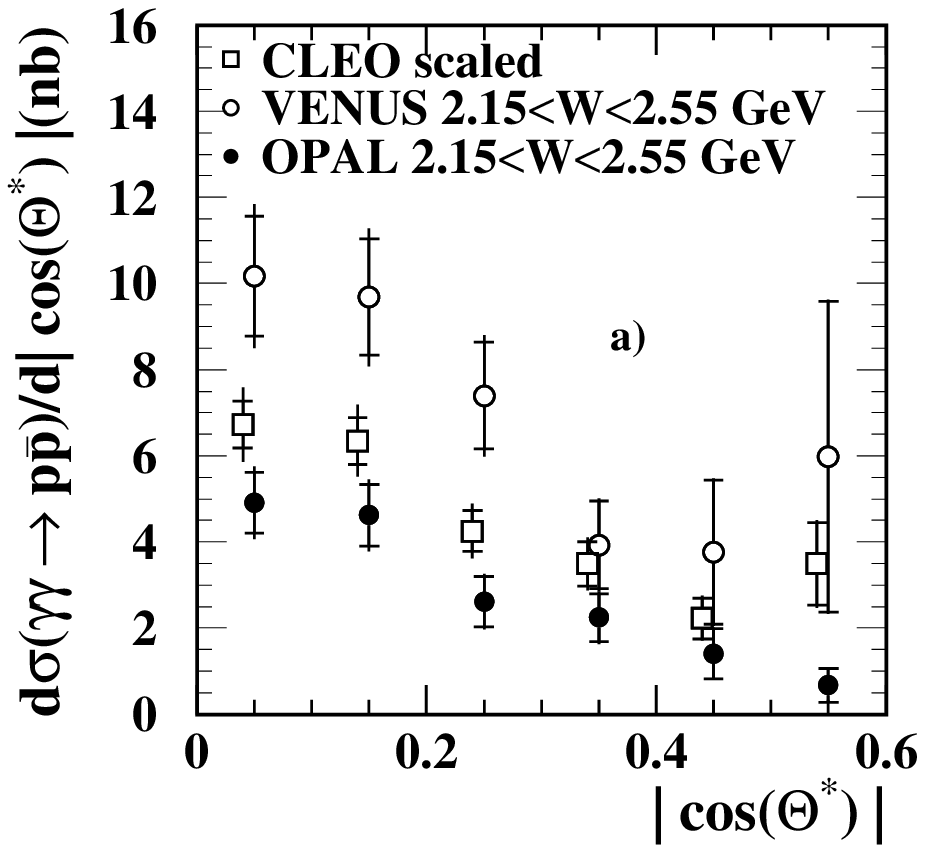}}
   \resizebox{0.48\textwidth}{!}{%
   \includegraphics{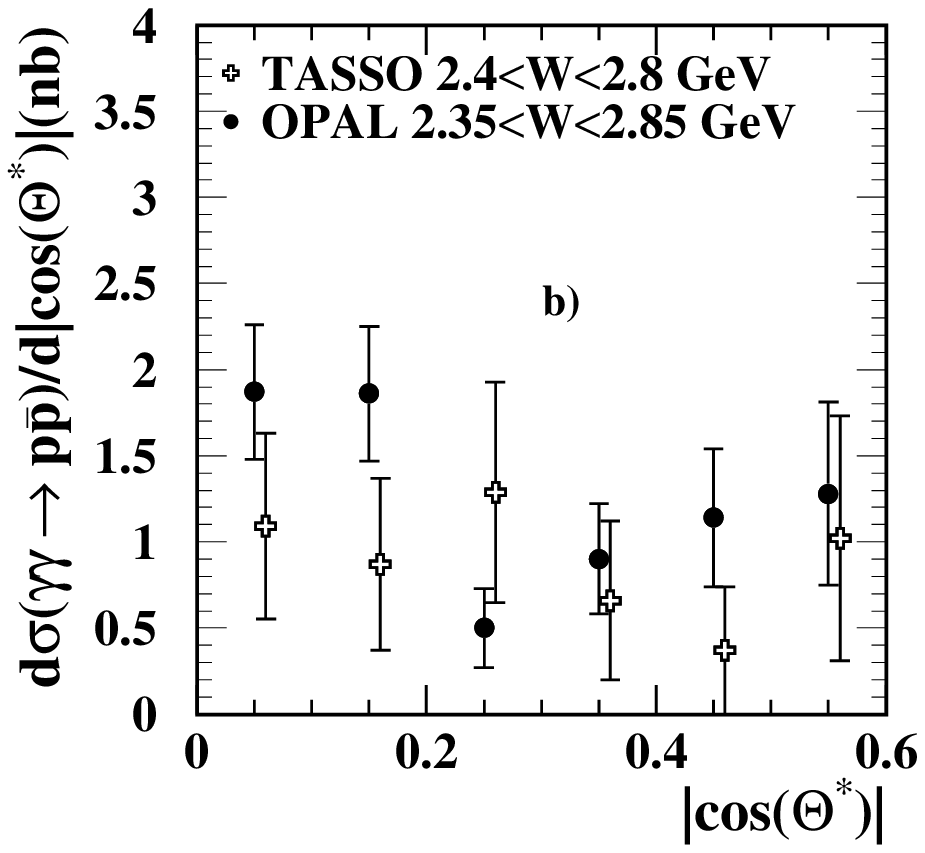}}
   \resizebox{0.48\textwidth}{!}{%
   \includegraphics{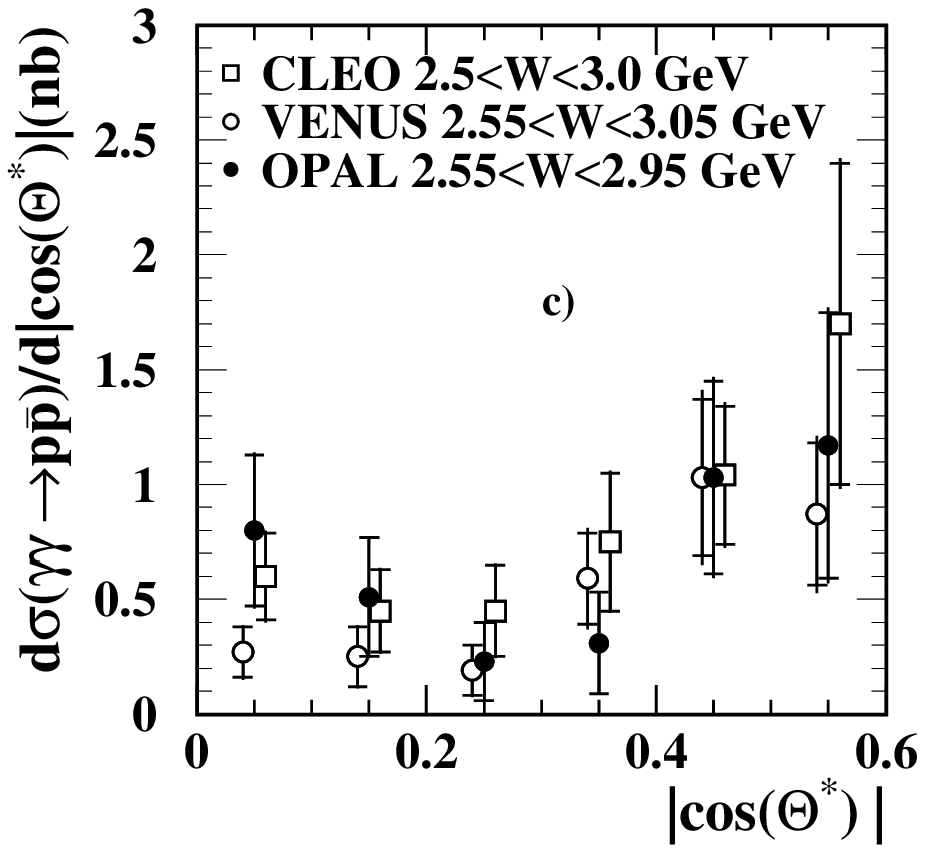}}
   \caption{Differential cross-sections for $\GG\to\PP$ as 
            a function of $\costs$ in different ranges of $W$; 
            a, c) compared with CLEO~\cite{Artuso:1994xk}
            and VENUS~\cite{Hamasaki:1997cy} data with statistical 
            (inner error bars) and systematic errors (outer bars), and
            b) compared with TASSO~\cite{Althoff:1983pf}.
            The TASSO error bars are statistical only. The data points
            are slightly displaced for clarity.} 
  \label{fig:cos1}
 \end{figure}
\end{center}

The measured differential cross-sections 
${{\rm d}\SI{(\GG\to\PP)}}/{{\rm d}\costs}$ in different $W$
ranges and for $\costs<0.6$ 
are given in Table~\ref{tab:cross59meas} and in Fig.~\ref{fig:cos1}.
The differential cross-section in the range $2.15<W<2.55\,\GV$  
lies below the results reported by 
VENUS~\cite{Hamasaki:1997cy} and CLEO~\cite{Artuso:1994xk} 
(Fig.~\ref{fig:cos1}a). 
Since the CLEO measurements are given for the lower $W$ range $2.0<W<2.5\,\GV$,
we rescale their results by a factor 0.635 which is the ratio of the two CLEO
total cross-section measurements integrated over the $W$ ranges
$2.0<W<2.5\,\GV$ and $2.15<W<2.55\,\GV$. This leads 
to a better agreement between the two measurements but the OPAL results
are still consistently lower.
The shapes of the $\costs$ dependence of all measurements are 
consistent apart from the highest $\costs$ bin, 
where the OPAL measurement is significantly lower than the measurements of
the other two experiments.

In Fig.~\ref{fig:cos1}b-c the differential cross-sections 
${{\rm d}\SI{(\GG\to\PP)}}/{{\rm d}\costs}$
in the $W$ ranges $2.35<W<2.85\,\GV$ and $2.55<W<2.95\,\GV$
are compared to the 
measurements by TASSO, VENUS and CLEO in similar $W$ ranges.
The measurements are consistent within the uncertainties. 

\begin{figure}[htbp]
 \centering
  \resizebox{0.6\textwidth}{!}{%
  \includegraphics{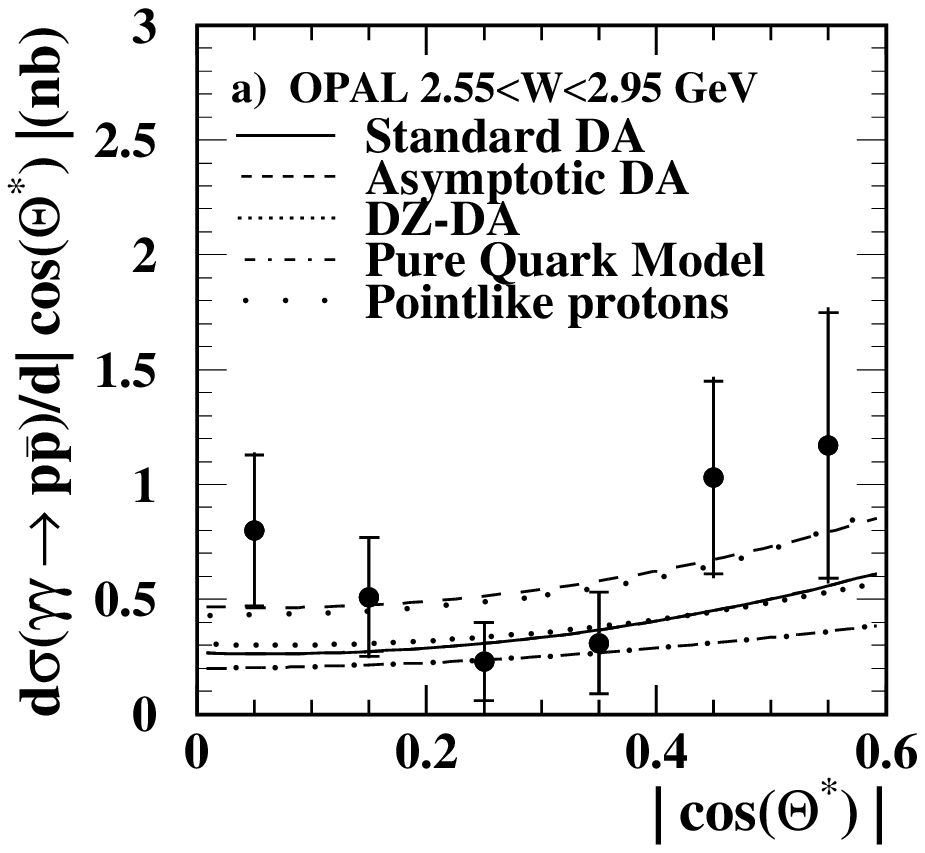}}
  \resizebox{0.6\textwidth}{!}{%
  \includegraphics{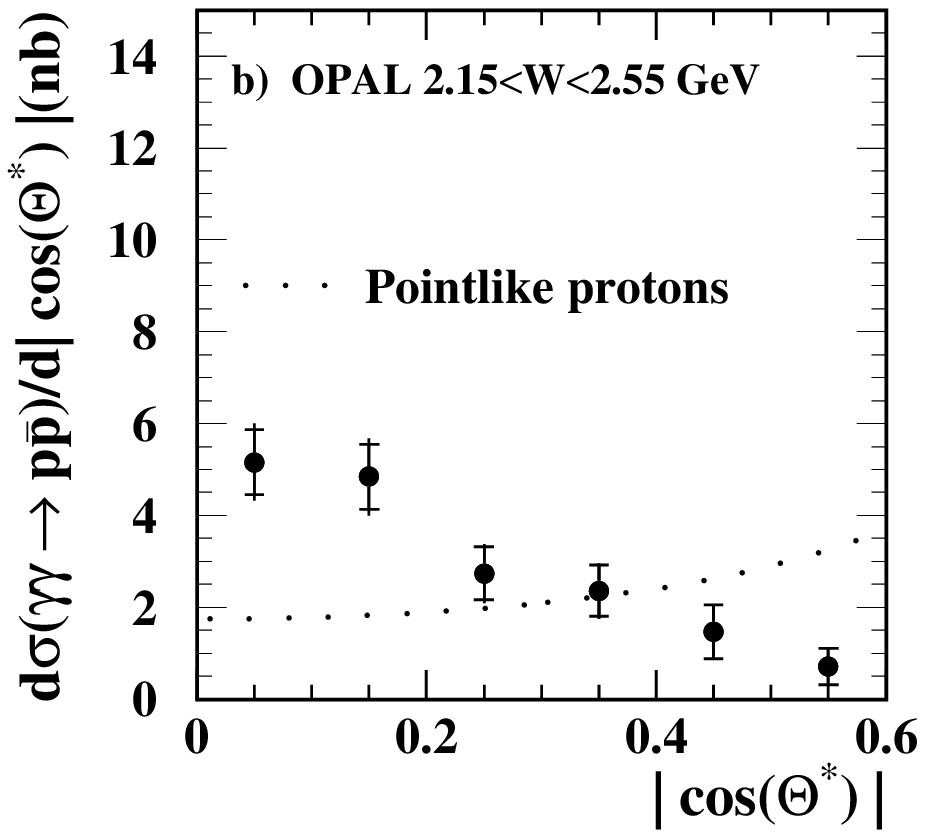}}
    \caption{Measured differential cross-section, 
      ${\rm d}\SI{(\GG\to\PP)}/{\rm d}\costs$, with statistical
      (inner bars) and total uncertainties (outer bars) for 
      a) $2.55<W<2.95\,\GV$ and 
      b) $2.15<W<2.55\,\GV$. 
      The data are compared with
      the point-like approximation for the proton  
      (\ref{eq:costest}) scaled to fit the data.
      The other curves show the 
      pure quark model~\protect\cite{Farrar:1985gv},
      the diquark model of\protect~\cite{Ansel:1987vk} with
      the Dziembowski distribution amplitudes (DZ-DA), and
      the diquark model of~\cite{berger:1997} using standard
      and asymptotic distribution amplitudes.}
    \label{fig:cos2}
\end{figure}

The comparison of the differential cross-section as a function 
of $\costs$ for $2.55<W<2.95\,\GV$ with the calculation 
of~\cite{berger:1997} at $W = 2.8\,\GV$ for different distribution 
amplitudes (DA) is shown in Fig.~\ref{fig:cos2}a.
The pure quark model~\cite{Farrar:1985gv,Millers:1986ca} 
and the quark-diquark model predictions 
lie below the data, but the shapes of the curves are consistent with 
those of the data. 

In Fig.~\ref{fig:cos2}b the differential cross-section 
${\rm d}\SI{(\GG\to\PP)}/{\rm d}\costs$ is shown versus $\costs$ 
for $2.15<W<2.55\,\GV$. 
The cross-section decreases at large $\costs$; the shape of the angular
distribution is different from that at higher $W$ values.
This indicates that for low $W$ the perturbative calculations
of~\cite{Farrar:1985gv,Millers:1986ca} are not valid.

Another important consequence of the hard scattering picture
is the hadron helicity
conservation rule. For each exclusive reaction like
$\GG\to\PP$ the sum of the two initial helicities equals
the sum of the two final ones~\cite{Brodsky:1981kj}.
According to the simplification used in~\cite{Ansel:1987vk}, 
only scalar diquarks are considered, and the (anti) proton carries the 
helicity of the single (anti) quark. 
Neglecting quark masses, quark and antiquark and hence proton and 
antiproton have to be in opposite helicity states. 
If the (anti) proton is considered as a point-like particle, simple 
QED rules determine the angular dependence of the unpolarized 
$\GG\to\PP$ differential cross-section~\cite{Budnev:1974de}: 
\begin{equation}
  \frac{{\rm d}\SI{(\GG\to\PP)}}{{\rm d}\costs} \propto \frac{(1 + \cos^{2}\theta^{*})}{(1 - \cos^{2}\theta^{*})}.
  \label{eq:costest}
\end{equation}
This expression is compared to the data in two $W$ ranges, $2.55<W<2.95\,\GV$ 
(Fig.~\ref{fig:cos2}a) and $2.15<W<2.55\,\GV$ (Fig.~\ref{fig:cos2}b). 
The normalisation in each case is determined by the best fit to 
the data. In the higher $W$ range, the prediction (\ref{eq:costest}) 
is in agreement with 
the data within the experimental uncertainties. 
In the lower $W$ range this 
simple model does not describe the data. At low $W$ soft 
processes such as meson exchange are expected to introduce other partial
waves, so that the approximations leading to (\ref{eq:costest})
become invalid~\cite{Brodsky:1987nt}.

\section{Conclusions}
The cross-section for the process $\EE\to\EE\PP$ has been measured  
in the $\PP$ centre-of-mass energy
range of $2.15 <W< 3.95\,\GV$ using data
taken with the OPAL detector at $\SQS = 183$ and $189\,\GV$.
The measurement extends to slightly larger values of $W$
than in previous measurements.

The total cross-section $\SI(\GG\to\PP)$ as a function of $W$  
is obtained from the differential cross-section
${\rm d}\SI(\EE\to\EE\PP)/{\rm d}W$ using a luminosity
function.
For the high $\PP$ centre-of-mass energies, $W>2.3\,\GV$, the measured 
cross-section is in good agreement 
with other experimental results~\cite{Althoff:1983pf,Aihara:1987ha,Albrecht:1989hz,Artuso:1994xk,Hamasaki:1997cy}.
At lower $W$ the OPAL measurements lie below the results
obtained by CLEO~\cite{Artuso:1994xk}, and VENUS~\cite{Hamasaki:1997cy},
but agree with the JADE~\cite{Bartel:1986sy} and 
ARGUS~\cite{Albrecht:1989hz} measurements.
The cross-section  as a function of $W$ is in agreement with
the quark-diquark model predictions of~\cite{Ansel:1987vk,berger:1997}.

The power law fit yields an exponent
$n=7.5\pm0.8$ where the uncertainty is statistical only. 
Within this uncertainty,
the measurement is not able to distinguish between predictions for 
the proton to interact as a state of three quasi-free quarks 
or as a quark-diquark system.
These predictions are based on dimensional counting 
rules~\cite{Brodsky:1973kr,Matveev:1973ra}. 
 
The shape of the differential cross-section 
${\rm d}\SI{(\GG\to\PP)}/{\rm d}\costs$
agrees with the results of previous experiments in comparable
$W$ ranges, apart from in the highest $\costs$ bin measured in
the range $2.15<W<2.55$~GeV. 
In this low $W$ region 
contributions from soft processes such as meson exchange are expected
to complicate the picture by introducing extra partial waves, and
the shape of the measured differential cross-section 
${\rm d}\SI{(\GG\to\PP)}/{\rm d}\costs$
does not agree with the simple model that leads to
the helicity conservation rule.
In the high $W$ region, $2.55<W<2.95\,\GV$, the 
experimental and theoretical differential cross-sections
${\rm d}\SI{(\GG\to\PP)}/{\rm d}\costs$
agree, indicating that the data are consistent with the
helicity conservation rule.
\section*{Acknowledgements}
We want to thank Mauro Anselmino, Wolfgang Schweiger,
Carola F. Berger, Maria Novella Kienzle-Focacci
and Sven Menke for important and fruitful discussions. 
We particularly wish to thank the SL Division for the efficient operation
of the LEP accelerator at all energies
 and for their close cooperation with
our experimental group.  In addition to the support staff at our own
institutions we are pleased to acknowledge the  \\
Department of Energy, USA, \\
National Science Foundation, USA, \\
Particle Physics and Astronomy Research Council, UK, \\
Natural Sciences and Engineering Research Council, Canada, \\
Israel Science Foundation, administered by the Israel
Academy of Science and Humanities, \\
Benoziyo Center for High Energy Physics,\\
Japanese Ministry of Education, Culture, Sports, Science and
Technology (MEXT) and a grant under the MEXT International
Science Research Program,\\
Japanese Society for the Promotion of Science (JSPS),\\
German Israeli Bi-national Science Foundation (GIF), \\
Bundesministerium f\"ur Bildung und Forschung, Germany, \\
National Research Council of Canada, \\
Hungarian Foundation for Scientific Research, OTKA T-029328, 
and T-038240,\\
Fund for Scientific Research, Flanders, F.W.O.-Vlaanderen, Belgium.\\

\newpage
\begin{table}[htbp]
  \begin{center}
   \begin{tabular}{|lccc|}
    \hline
     Monte Carlo                   &Number of events &Luminosity       & upper limit \\
     background process            &generated        & (fb$^{-1}$)     & at $95\,\%$ CL      \\
      \hline                             
$\EE\to\EE\EE$~\protect\cite{Vermaseren:1983cz}  &$800\,000$  & $\pz1.0 $& $<0.75$ \\
$\EE\to\EE\MuMu$~\protect\cite{Vermaseren:1983cz}&$600\,000$  & $\pz1.0 $& $<0.75$ \\ 
$\EE\to\EE\TT$~\protect\cite{Vermaseren:1983cz}  &$430\,000$  & $\pz1.0 $& $<0.76$ \\
$\EE\to\EE\PiPi$~\protect\cite{linde}           &$149\,000$  & $\pz0.6 $& $<1.29$ \\
$\EE\to\TT$~\protect\cite{Jadach:1994yv}         &$\pz80\,000$& $\pz1.0 $& $<0.74$ \\
$\EE\to\MuMu$~\protect\cite{Jadach:1994yv}       &$\pz80\,000$& $10.0$   & $<0.07$ \\
$\EE\to\EE$~\protect\cite{Jadach:1997nk}         &$600\,000$  & $\pz1.0 $& $<0.74$ \\
    \hline 
\end{tabular}
  \caption{Number of generated events and corresponding integrated 
           luminosities for the different background processes. 
           Since no events pass the selection cuts,
           $95\%$ confidence level (CL) upper limits are given 
           for the number of background events expected to contribute 
           to the selected data sample. 
        } 
   \label{tab:mcbg}
 \end{center}
\end{table}
\begin{table}[htbp]
 \begin{center}
   \begin{tabular}{|c|l l|}
   \hline 
   trigger class               & data (\%) &  model (\%)       \\
  \hline 
  (2,1)                      & \Tritod   & \Tritofm    \\
  (2,0)                      & \Tritzd   & \Tritzfm    \\
  (1,1)                      & \Triood   & \Trioofm    \\
  (2,2)                      & \Trittd   & \Trittfm    \\\hline
  \end{tabular}
  \caption{Measured and predicted relative frequencies of four 
    different sub-combinations, 
    ({\ensuremath{N_{\rm TT}}},{\ensuremath{N_{\rm TOF}}}) 
    = (2,1), (2,0), (1,1) and
    (2,2), which can trigger the $\EE\to\EE\PP$ event for events with
    $W > 2.15\,\GV$. The uncertainties are statistical only. 
    }
  \label{tab:fractions}
 \end{center}
\end{table}
\begin{table}[htbp]        
 \begin{center}
  \begin{tabular}{|c|c|c|c|c|}
  \hline
  $W$ range &$\langle W\rangle$&Events &${\rm d}\SI(\EE\to\EE\PP)/{\rm d}W$& $\SI{(\GG\to\PP)}$ \\
     ($\GV$)    & ($\GV$)          &       & (pb/GeV)&  (nb)    \\ 
    \hline

    \Wab-\Wac & \MWb       & \Nab       &\Seeb  & \Sneb  \\
    \Wac-\Wad & \MWc       & \Nac       &\Seec  & \Snec  \\
    \Wad-\Wae & \MWd       & \Nad       &\Seed  & \Sned  \\
    \Wae-\Waf & \MWe       & \Nae       &\Seee  & \Snee  \\
    \Waf-\Wah & \MWfg      & \Nafg      &\Seefg & \Snefg \\
    \Wah-\Wam & \MWhl      & \Nahl      &\Seehl & \Snehl \\
    \Wam-\War & \MWlo      &\Namo       &\Seemo & \Snemo \\
    \War-\Waz & \MWpv      &\Napv       &\Seepv & \Snepv \\ 
    \hline
   \end{tabular}
   \caption{
          Number of events and cross-sections
          for $\costs<0.6$. Statistical and systematic uncertainties
          are also given.
          A Poisson asymmetric statistical uncertainty, determined within a 
          $68\%$ confidence interval, has been calculated for the last
          two bins of $W$ with only one selected event.
          } 
           \label{tab:crossmeas}
\end{center}
\end{table}
\begin{table}[htbp]
\vspace*{1.0cm}
 \begin{center}
  \begin{tabular}{|c|c|c|c|c|c|c|}
  \hline 
   $\costs$&\multicolumn{2}{c|}{$2.15<W<2.55\,\GV$}
           &\multicolumn{2}{c|}{$2.35<W<2.85\,\GV$}
           &\multicolumn{2}{c|}{$2.55<W<2.95\,\GV$}\\ \cline{2-7}  
  &$N_{\rm ev}$ 
  &$\frac {{\rm d}\sigma(\GG\to\PP)}{{\rm d}\costs}(\nb)$
  &$N_{\rm ev}$ 
  &$\frac {{\rm d}\sigma(\GG\to\PP)}{{\rm d}\costs}(\nb)$
  &$N_{\rm ev}$
  &$\frac {{\rm d}\sigma(\GG\to\PP)}{{\rm d}\costs}(\nb)$\\
   \hline 
   \bina        &\neaaa &\cstaaa    &\neab &\cstab &\nea  &   \csta  \\
   \binb        &\nebaa &\cstbaa    &\nebb &\cstbb &\neb  &   \cstb  \\
   \binc        &\necaa &\cstcaa    &\necb &\cstcb &\nec  &   \cstc  \\
   \bind        &\nedaa &\cstdaa    &\nedb &\cstdb &\ned  &   \cstd  \\
   \bine        &\neeaa &\csteaa    &\neeb &\csteb &\nee  &   \cste  \\
   \binf        &\nefaa &\cstfaa    &\nefb &\cstfb &\nef  &   \cstf  \\
   \hline
  \end{tabular}
  \caption{
  Number of data events selected in each 
  bin of $\costs$ and the measured differential cross-sections  
  ${\rm d}\sigma(\GG\to\PP)/{\rm d}\costs$
  for three $W$ ranges. 
  The statistical and systematic uncertainties are also given.
  } 
  \label{tab:cross59meas}
 \end{center}
\end{table}


\begin{table}
 \begin{center}
  \begin{tabular}{|l|c|}
  \hline 
   Source of Systematic                  & Systematic   \\
   uncertainties                         & uncertainty (\%)  \\
  \hline
   Luminosity Function                    & 5.0          \\ 
   Trigger Efficiency                     & 5.0          \\
   Monte Carlo statistics ($W<2.55$~GeV)   & 4.5          \\
   \hspace*{4.0cm}        ($W>2.55$~GeV)   & 6.0          \\
   $\DEDX$ cuts \hspace*{0.1cm} ($W<2.55$~GeV)   & 0.1    \\
   \hspace*{2.35cm}        ($W>2.55$~GeV)  & 5.0          \\
   Residual Background                    & 6.0          \\
  \hline
   Total  ($W<2.55$~GeV)                   & 10.3         \\
   Total  ($W>2.55$~GeV)                   & 12.1         \\
  \hline
  \end{tabular}
 \caption{Estimates of relative systematic uncertainties. The total systematic 
     uncertainty is obtained by adding them in quadrature.
     }          
 \label{tab:sys}
 \end{center}
\end{table}

\end{document}